\documentclass[reprint,aps,pra,amsmath,amssymb,amsfonts,floatfix,superscriptaddress]{revtex4-2}

\usepackage{graphicx}
\usepackage{color}
\usepackage{dcolumn}
\usepackage{bm}
\usepackage{hyperref}
\usepackage[utf8]{inputenc}
\usepackage[normalem]{ulem}

\newcommand{\ket}[1]{\ensuremath{\left|{#1}\right\rangle}}
\newcommand{\C}[2]{\parbox[t]{#1\columnwidth}{\raggedright #2\strut}}

\begin{document}

\title{Stimulated Parametric Down-Conversion: From Foundational Coherence to Structured Light Applications}

\author{M. F. Z. Arruda}
\email{marcelo.arruda@ifmt.edu.br}
\affiliation{Instituto Federal de Mato Grosso, CEP 78895-150, Sorriso, MT, Brazil}

\author{M. G. Damaceno}
\affiliation{Departamento de F\'{i}sica, Universidade Federal de Santa Catarina, CEP 88040-900, Florian\'{o}polis, SC, Brazil}

\author{G. H. dos Santos}
\email{fisica.gu@gmail.com}
\affiliation{Departamento de F\'{\i}sica, Universidad de Concepci\'on, 160-C Concepci\'on, Chile}
\affiliation{Millennium Institute for Research in Optics, Universidad de Concepci\'on, 160-C Concepci\'on, Chile}

\author{R. Medeiros de Araújo}
\email{renne.araujo@ufsc.br}
\affiliation{Departamento de F\'{i}sica, Universidade Federal de Santa Catarina, CEP 88040-900, Florian\'{o}polis, SC, Brazil}

\author{N. Rubiano da Silva}
\email{nara.rubiano@ufsc.br}
\affiliation{Departamento de F\'{i}sica, Universidade Federal de Santa Catarina, CEP 88040-900, Florian\'{o}polis, SC, Brazil}

\author{S. P. Walborn}
\email{swalborn@udec.cl}
\affiliation{Departamento de F\'{\i}sica, Universidad de Concepci\'on, 160-C Concepci\'on, Chile}
\affiliation{Millennium Institute for Research in Optics, Universidad de Concepci\'on, 160-C Concepci\'on, Chile}

\author{P. H. Souto Ribeiro}
\email{p.h.s.ribeiro@ufsc.br}
\affiliation{Departamento de F\'{i}sica, Universidade Federal de Santa Catarina, CEP 88040-900, Florian\'{o}polis, SC, Brazil}

\date{\today}

\begin{abstract}
Stimulated parametric down-conversion (StimPDC) is the seeded version of the
process that underlies much of photonic quantum information. Replacing the vacuum
at one input port with a coherent seed leaves the interaction Hamiltonian, the
phase-matching function and the two-photon amplitude unchanged, while raising the
output to levels accessible with ordinary classical detectors. This review treats
the transverse spatial physics of that process. We show that a single relation, in
which the idler amplitude is the pointwise product of the pump field with the
complex conjugate of the seed, organizes results usually presented separately: the
image and coherence transfer experiments of the 1990s; phase conjugation and its
use for wavefront correction; the conservation and conjugation of orbital angular
momentum, together with the generation of radial modes and geometric intensity
patterns; and the processing of vector vortex beams carrying spatially varying
polarization. We then consider StimPDC as an instrument rather than as an object
of study, covering stimulated emission tomography, the experimental realization of
Klyshko's advanced-wave picture, and the transfer of partial coherence, including
the conservation of the twist phase of twisted Gaussian Schell-model beams. We
close with the quantum regime, reached when the seed approaches the single-photon
level or the idler is used as a herald, and with an assessment of open problems
and prospective applications.
\end{abstract}

\maketitle

\section{Introduction}
\label{sec:intro}
Parametric down-conversion (PDC) is a cornerstone of nonlinear and quantum
optics. In its spontaneous form (SPDC), vacuum fluctuations seed a
second-order ($\chi^{(2)}$) interaction that converts pump photons into
correlated signal and idler photon pairs, a standard resource for
entanglement-based quantum information. When the vacuum at the signal or idler
port is replaced by a coherent ``seed'' laser, the same interaction proceeds
in its stimulated form (StimPDC)~\cite{ou90}. 
The foundations of the field were laid in studies of spatial coherence and
image transfer. Early experiments established that the visibility of
interference, the angular spectrum, and the coherence properties of the pump
and seed are transferred to the down-converted field, culminating in the
direct observation of image transfer and phase conjugation in
StimPDC~\cite{Ribeiro95,monken1998transfer,fonseca1999quantum,souto2001observation}.
These results fixed the basic picture: StimPDC is a coherent, mode-selective
process that imprints the structure of the input fields onto the output. The consequences go well beyond
an increase in brightness: because the seed fixes the mode into which
down-conversion is stimulated, the spatial,
spectral, and polarization degrees of freedom are mapped deterministically
onto the generated field, and the output is raised to levels accessible with
ordinary classical detectors~\cite{Ribeiro95}.
These two properties, mode selectivity and classical brightness, organize both the landscape of StimPDC research and applications, as well as this review. 

That the
stimulated idler is generated as a coherent contraction of the seed against
the two-photon amplitude of the source leads to two main consequences. First, if the signal and idler are
maximally entangled in the SPDC regime, the idler in StimPDC is an exact,
phase conjugate of the seed~\cite{souto2001observation}. This conjugation relation, together with the rotational invariance of the
three-wave-mixing Hamiltonian that enforces conservation of orbital angular
momentum (OAM)~\cite{caetano2002conservation,Oliveira21}, accounts for the
generation of geometric patterns through nonlinear mixing of optical
vortices~\cite{Santos2022}, and for the phase conjugation of vector vortex beams
that carry spatially varying
polarization~\cite{deoliveira2020realtime,deoliveira2020quantum,daSilva2021PRA}.
Recent work has continued to
expand the catalog of spatial modes that StimPDC can synthesize and
tailor~\cite{g22lcgr1}. This has been used for orthogonal spatial coding
and nonlocal aberration cancellation~\cite{Xu:23} and, more recently, for
all-optical turbulence mitigation in free-space quantum key
distribution~\cite{aguilarcardoso2025turb}. 

Second, the seed beam can be used to probe the two-photon SPDC amplitude, making low-gain StimPDC a useful
characterization tool. Stimulated emission tomography (SET) exploits the fact
that a bright, classically detectable stimulated signal carries the same
information as a photon-coincidence measurement of the spontaneous pair, with
count rates larger by up to nine orders of magnitude~\cite{Liscidini2013}.
First demonstrated in the spectral and polarization degrees of
freedom~\cite{eckstein2014,Rozema:15,Fang:16,Ciampini:19} and placed on a
complete modal footing in the low-gain regime~\cite{Keller2022}, SET has more
recently been applied to the transverse spatial degree of freedom,
reconstructing the joint spatial mode distribution of an entangled
pair~\cite{Xu2024}. 

Finally, StimPDC reaches into the quantum regime when the seed is pushed toward
the single-photon level or used to drive elementary quantum operations. A
heralded, cascaded scheme can realize single-photon OAM conjugation despite the
low flux of true single-photon seeds~\cite{Luck26}. Stimulated emission in a
parametric amplifier realizes optimal quantum cloning and the universal-NOT
gate~\cite{lamaslinares2002,demartini2002unot}, the latter being the
single-qubit operation that the vector phase conjugation noted above realizes
for one of the Bell-state source configurations~\cite{deoliveira2020quantum,daSilva2021PRA}. Detection of the idler photon can be used as a herald for 
conditional photon addition on the seed field. This approach has produced
single-photon-added coherent states and single-photon added thermal states that interpolate between the quantum and
classical descriptions of light~\cite{zavatta2004,parigi2007,kiesel2007,kiesel2011}. In this direction is the closely related field of imaging with undetected photons and induced
coherence, and adjacent quantum-regime use of parametric sources~\cite{ZouWangMandel1991,Lemos2014,Defienne2024}.

This review surveys the characterization and applications of
StimPDC, taking the transverse spatial degree of freedom and structured light as its focal point. The remainder of the paper is organized as follows.
Section~\ref{sec:theory} develops the idler-field amplitude and the
phase-conjugation relation. Sections~\ref{sec:fundamentals}--\ref{sec:vvb} review spatial
coherence and image transfer, structured light and OAM, and vector vortex
beams. Section~\ref{sec:nonlinear} covers stimulated emission tomography as a
characterization paradigm, nonlinear imaging, holographic
(advanced-wave) modeling, and partially coherent beams.
Section~\ref{sec:quantum} surveys StimPDC in the
quantum regime. Section~\ref{sec:outlook} discusses applications and outlook
before we conclude.

\section{Theory}
\label{sec:theory}

\subsection{Idler Field Amplitude in StimPDC}
\label{subsec:idlerStimPDC}

\begin{figure}[t]
\centering
\includegraphics[width=\linewidth]{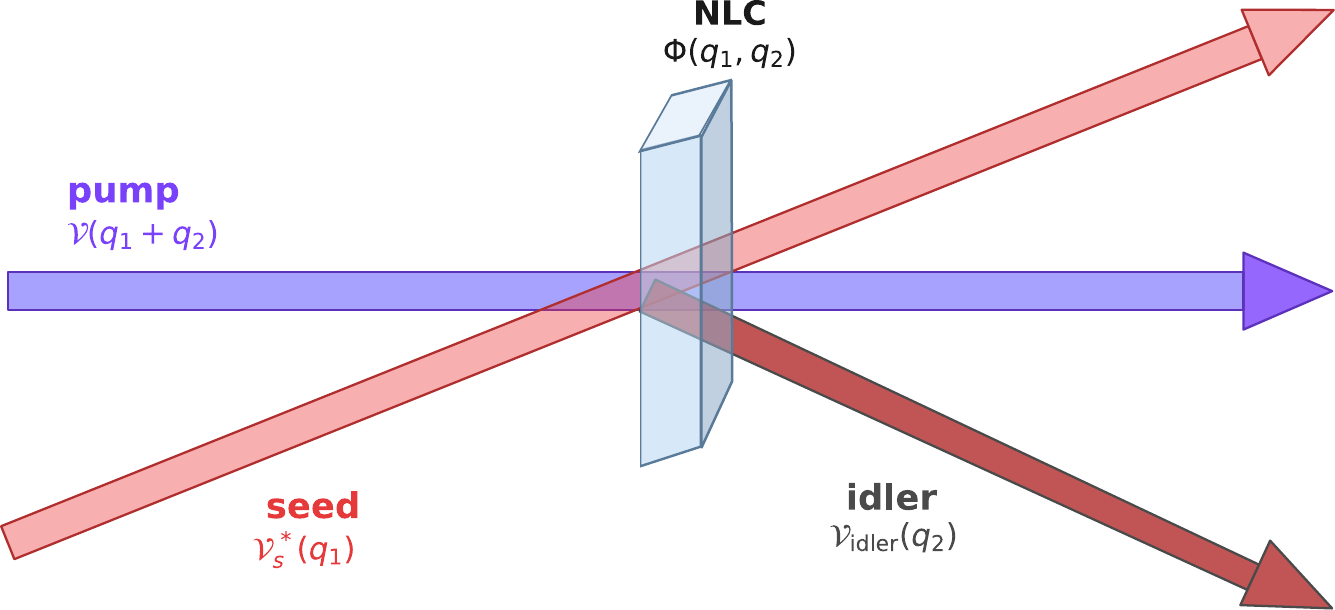}
\caption{Typical StimPDC experimental scheme. NLC is the nonlinear crystal, with $\Phi(\mathbf q_1,\mathbf q_2)$ being the function describing the conversion of the pump field $v(\mathbf q_1+\mathbf q_2)$ by the NLC. The functions $v_s^*(\mathbf q_1)$ and $v_\text{idler}(\mathbf q_2)$ are the seed and idler fields, respectively. All functions relate to the corresponding angular spectra. See more details in the main text.}
\label{fig1}
\end{figure}

The theoretical framework of StimPDC is rooted in the nonlinear interaction of three optical fields within a second order ($\chi^{(2)}$) nonlinear medium. While spontaneous PDC is initiated by vacuum fluctuations, the presence of an
external ``seed'' field in the stimulated regime changes the emission statistics
qualitatively: down-conversion is amplified into the mode occupied by the seed
and the output acquires the coherence of the input fields. It is worth
emphasizing at the outset that this does \emph{not} require large parametric
gain. Throughout this review we work in the undepleted-pump, low-gain regime, up to
first order in the interaction Hamiltonian, so that the enhancement of the idler
is proportional to the mean photon number $|\alpha|^{2}$ of the seed rather than
to an exponential gain factor. The distinct high-gain regime of parametric
down-conversion, in which the output is a bright squeezed vacuum and higher
orders of the Bogoliubov transformation must be retained, lies outside our scope, though the production of entangled images in the high-gain regime is briefly mentioned in Sec.~\ref{subsec:highgain}. 
We note also that Ref.~\cite{Chekhova2016NLI} provides a review of the high-gain regime and of the
closely related SU(1,1) nonlinear interferometers. In Fig.~\ref{fig1}, the typical experimental setup is shown.

In the low-gain regime considered here, StimPDC can often be macroscopically described via classical optics. However, because it continuously transitions to the nonclassical SPDC regime as the seed field approaches the vacuum state and since both processes share the same underlying mode coupling mechanisms, it is highly advantageous to describe StimPDC using the same quantum mechanical formalism as SPDC. This quantum approach was originally pioneered by Wang and Mandel to obtain a quantum description of the time-frequency degrees of freedom in stimulated processes~\cite{Wang1994}. It was subsequently adapted by Monken et al. to describe the transverse spatial degrees of freedom in SPDC~\cite{monken1998transfer}, and then to StimPDC~\cite{ribeiro1999}, yielding the state vector:
\begin{align}
\ket{\psi}
=&
\ket{\alpha;v_s}_1\ket{0}_2
+
C
\iint
d\mathbf q_1
d\mathbf q_2\,
v(\mathbf q_1+\mathbf q_2)
\gamma(\mathbf q_1-\mathbf q_2)
\nonumber\\
\times\,
&
\hat a_1^\dagger(\mathbf q_1)
\ket{\alpha;v_s}_1
\otimes
\ket{1;\mathbf q_2}_2.
\label{eq:StimPDC_state}
\end{align}
Here, $\ket{1;\mathbf q_2}_2$ denotes a single photon in the plane-wave mode with transverse momentum $\mathbf q_2$, $v$ is the pump angular spectrum and $\gamma$ the phase-matching function. The ket $\ket{\alpha;v_s}_1$ denotes a multimode coherent state in the signal
channel (mode 1) whose photons are distributed over transverse wavevectors
according to a normalized angular spectrum $v_s$,
with total complex amplitude $\alpha$. 
It is generated by the multimode displacement operator
\begin{equation}
\ket{\alpha;v_s}_1 = \hat D_1[\alpha v_s]\,\ket{0}_1,
\label{eq:displacement}
\end{equation}
with $\hat D_1[\alpha v_s] = e^{\hat S}$
with the generator
$\hat S \equiv \alpha\,\hat A_1^\dagger(v_s) - \alpha^{*}\hat A_1(v_s)$ and the multimode creation operator defined as
\begin{equation}
\hat A_1^\dagger(v_s) \equiv \int d\mathbf q\; v_s(\mathbf q)\,
\hat a_1^\dagger(\mathbf q),
\label{eq:collective}
\end{equation}
which creates one photon in the single transverse mode whose profile is $v_s$. 
We note now that the commutator of a
single plane-wave annihilation operator with the generator $\hat S$ is a $c$-number,
\begin{align}
\left[\hat a_1(\mathbf q),\hat S\right]
&= \alpha \int d\mathbf q'\; v_s(\mathbf q')
\left[\hat a_1(\mathbf q),\hat a_1^\dagger(\mathbf q')\right]
\nonumber\\
&= \alpha \int d\mathbf q'\; v_s(\mathbf q')\,
\delta(\mathbf q-\mathbf q')
\;=\; \alpha\, v_s(\mathbf q),
\label{eq:commutator_cnumber}
\end{align}
which leads to 
\begin{equation}
\hat D_1^\dagger[\alpha v_s]\;\hat a_1(\mathbf q)\;\hat D_1[\alpha v_s]
= \hat a_1(\mathbf q) + \alpha\, v_s(\mathbf q).
\label{eq:displaced_operator}
\end{equation}
Then,
\begin{align}
\hat a_1(\mathbf q)\ket{\alpha;v_s}_1
&= \hat D_1 \left(\hat D_1^\dagger \hat a_1(\mathbf q) \hat D_1\right)\ket{0}_1
\nonumber\\
&= \alpha\, v_s(\mathbf q)\,\ket{\alpha;v_s}_1 .
\label{eq:eigenvalue}
\end{align} 
$|\alpha|^{2}$ is the mean photon number of the seed occupying the mode with transverse momentum $\mathbf q$. In Sec.~\ref{sec:fundamentals} it is related with the occupation number per mode measured in the
earliest StimPDC experiments~\cite{Ribeiro95}. 
\par
The spatial intensity distribution of the generated idler beam (mode 2) is given by the expectation value:
\begin{equation}
\mathcal I(\boldsymbol{\rho}_2)
=
\left\langle
\hat E_2^{(-)}(\boldsymbol{\rho}_2)
\hat E_2^{(+)}(\boldsymbol{\rho}_2)
\right\rangle.
\label{eq:idler_intensity}
\end{equation}
Using the positive-frequency component of the electric field operator in the paraxial approximation,
\begin{equation}
\hat E_2^{(+)}(\boldsymbol{\rho}_2)
=
\frac{1}{2\pi}
\int
d\mathbf q_2' \,
\hat a_2(\mathbf q_2') \,
e^{i\mathbf q_2'\cdot\boldsymbol{\rho}_2},
\label{eq:stim_field}
\end{equation}
the action of the annihilation operator on the state vector $\ket{\psi}$ yields:
\begin{align}
\hat E_2^{(+)}(\boldsymbol{\rho}_2)
\ket{\psi}
\propto&
\iint
d\mathbf q_1
d\mathbf q_2 \,
e^{i\mathbf q_2\cdot\boldsymbol{\rho}_2}
\,
\Phi(\mathbf q_1,\mathbf q_2)
\hat a_1^\dagger(\mathbf q_1)
\ket{\alpha;v_s}_1
\ket{0}_2,
\label{eq:Epsi}
\end{align}
where we define the two-photon angular spectrum
\begin{equation}
 \Phi(\mathbf q_1,\mathbf q_2) = v(\mathbf q_1+\mathbf q_2)\gamma(\mathbf q_1-\mathbf q_2).
 \label{eq:Phi}
\end{equation}

Substituting Eq.~\eqref{eq:Epsi} into Eq.~\eqref{eq:idler_intensity}, the explicit expression for the idler intensity becomes:
\begin{align}
\mathcal I(\boldsymbol{\rho}_2)
\propto
\int
d\mathbf q_1
d\mathbf q_1'
d\mathbf q_2
d\mathbf q_2'
\,
&
e^{i(\mathbf q_2-\mathbf q_2')\cdot\boldsymbol{\rho}_2}
\,
\Phi^*(\mathbf q_1',\mathbf q_2')
\Phi(\mathbf q_1,\mathbf q_2)
\nonumber\\
\times\,
&
\left\langle
\alpha;v_s
\left|
\hat a_1(\mathbf q_1')
\hat a_1^\dagger(\mathbf q_1)
\right|
\alpha;v_s
\right\rangle.
\label{eq:general_intensity}
\end{align}
By applying the standard bosonic commutation relation,
\begin{equation}
[
\hat a_1(\mathbf q_1'),
\hat a_1^\dagger(\mathbf q_1)
]
=
\delta(\mathbf q_1-\mathbf q_1'),
\end{equation}
the coherent state expectation value expands to:
\begin{align}
\left\langle
\alpha;v_s
\left|
\hat a_1(\mathbf q_1')
\hat a_1^\dagger(\mathbf q_1)
\right|
\alpha;v_s
\right\rangle
=
\delta(\mathbf q_1-\mathbf q_1')
+
|\alpha|^2 v_s(\mathbf q_1')
v_s^*(\mathbf q_1).
\label{eq:coherent_average}
\end{align}
This evaluation naturally splits the total measured idler intensity into a sum of spontaneous and stimulated contributions:
\begin{equation}
\mathcal I(\boldsymbol{\rho}_2)
=
\mathcal I_{\mathrm{spont}}(\boldsymbol{\rho}_2)
+
\mathcal I_{\mathrm{stim}}(\boldsymbol{\rho}_2).
\label{eq:intensity_split}
\end{equation}
The $\delta$-function contribution to Eq.~\eqref{eq:coherent_average} is the
spontaneous term,
\begin{equation}
\mathcal I_{\mathrm{spont}}(\boldsymbol{\rho}_2)
\propto
\int d\mathbf q_1
\left|
\int d\mathbf q_2\,
e^{i\mathbf q_2\cdot\boldsymbol{\rho}_2}\,
\Phi(\mathbf q_1,\mathbf q_2)
\right|^{2},
\label{eq:spont_term}
\end{equation}
an \emph{incoherent} superposition over the signal modes $\mathbf q_1$: each
signal plane wave contributes its own idler amplitude and the contributions add
in intensity. The term proportional to $v_s(\mathbf q_1')v_s^{*}(\mathbf q_1)$,
by contrast, factorizes into a single squared modulus and is therefore
\emph{coherent}:
\begin{align}
\mathcal I_{\mathrm{stim}}(\boldsymbol{\rho}_2)
\propto
|\alpha|^2 \Bigg|
\iint
d\mathbf q_1
d\mathbf q_2
\,
e^{i\mathbf q_2\cdot\boldsymbol{\rho}_2}
\,
\Phi(\mathbf q_1,\mathbf q_2)
v_s^*(\mathbf q_1)
\Bigg|^2.
\label{eq:stimulated_term}
\end{align}
This is the structural origin of every coherence effect
reviewed in Sec.~\ref{sec:fundamentals}: the seed does not merely brighten the
idler, it replaces an incoherent mode sum by a coherent one, and the ratio
$\mathcal I_{\mathrm{stim}}/\mathcal I_{\mathrm{spont}}$ is precisely
the occupation number per mode that controls the observed fringe
visibility~\cite{Ribeiro95}.
For a bright seed, $|\alpha|^{2}\gg 1$, the stimulated contribution is typically
several orders of magnitude stronger than the spontaneous background and
dominates the profile 
$\mathcal I(\boldsymbol{\rho}_2) \sim \mathcal I_{\mathrm{stim}}(\boldsymbol{\rho}_2)$.

The effective classical electric field amplitude of the stimulated idler beam (omitting a global phase factor) is given by:
\begin{equation}
\mathcal A_{\mathrm{stim}}(\boldsymbol{\rho}_2)
\propto \alpha
\iint
d\mathbf q_1
d\mathbf q_2
\,
e^{i\mathbf q_2\cdot\boldsymbol{\rho}_2}
\,
\Phi(\mathbf q_1,\mathbf q_2)
v_s^*(\mathbf q_1).
\label{eq:stimulated_term_amp}
\end{equation}
This result demonstrates that the generated idler field in StimPDC inherits a structural dependence on both the pump envelope and the phase matching properties, mirroring the behavior observed in conventional SPDC. Crucially, however, the stimulated configuration introduces an additional, decisive layer of control through its direct dependence on the spatial and coherence profiles of the external seed field. We include the coherent amplitude $\alpha$ in the above equation to emphasize its presence with regard to the SPDC amplitude. It will be excluded when not necessary.

\subsection{Angular Spectrum and Image Transfer}
\label{subsec:asit}
Let us write the idler spatial field amplitude $\mathcal A_{\mathrm{stim}}(\boldsymbol{\rho}_2)$ as the inverse Fourier transform of its angular spectrum $v_\text{idler}(\mathbf q_2)$:
\begin{equation}
\mathcal A_{\mathrm{stim}}(\boldsymbol{\rho}_2) \propto \int d\mathbf q_2 \, e^{i\mathbf q_2\cdot\boldsymbol{\rho}_2} \, v_\text{idler}(\mathbf q_2).
\end{equation}
By directly comparing the two integrals, we can isolate the angular spectrum of the idler field:
\begin{equation}\label{eq:idler_field}
v_\text{idler}(\mathbf q_2) \propto \int d\mathbf q_1 \, v(\mathbf q_1+\mathbf q_2) \gamma(\mathbf q_1-\mathbf q_2) v_s^*(\mathbf q_1),
\end{equation}
where we used Eq.~\eqref{eq:Phi}. The idler field thus depends upon three independent functions.

\subsubsection{The thin crystal limit}
\label{subsec:product}
Equations~\eqref{eq:stimulated_term_amp} and~\eqref{eq:idler_field} already contain all of the physics
reviewed in this paper, but its content becomes even more transparent only after it is
translated to position space. In this regard, let us adopt the thin-crystal approximation,
$\gamma(\mathbf q_1-\mathbf q_2)\rightarrow 1$, which applies to most SPDC scenarios~\cite{walborn2010}. This limit will be lifted below, but for the time being it assists in identifying several interesting scenarios in StimPDC. We can then make a change of variables
$\mathbf Q = \mathbf q_1+\mathbf q_2$ to decouple the two integrals in~\eqref{eq:stimulated_term_amp},
\begin{align}
\mathcal A_{\mathrm{stim}}(\boldsymbol\rho)
&\propto
\iint d\mathbf Q\, d\mathbf q_1\;
e^{i(\mathbf Q-\mathbf q_1)\cdot\boldsymbol\rho}\;
v(\mathbf Q)\, v_s^{*}(\mathbf q_1)
\nonumber\\[2pt]
&=
\left[\int d\mathbf Q\; v(\mathbf Q)\, e^{i\mathbf Q\cdot\boldsymbol\rho}\right]
\left[\int d\mathbf q_1\; v_s(\mathbf q_1)\, e^{i\mathbf q_1\cdot\boldsymbol\rho}\right]^{*},
\end{align}
so that, up to a constant,
\begin{equation}
\mathcal A_{\mathrm{stim}}(\boldsymbol\rho)
\;\propto\;
\mathcal E_p(\boldsymbol\rho)\,\mathcal E_s^{*}(\boldsymbol\rho),
\label{eq:product}
\end{equation}
where $\mathcal E_p$ and $\mathcal E_s$ are the transverse profiles of
the pump and seed at the crystal plane. We will refer to
Eq.~\eqref{eq:product} as the \emph{amplitude product relation}. It states that StimPDC
performs a pointwise multiplication of the pump field by the conjugate of the
seed field, a spatial cross modulation of one beam by the other.

Away from the thin-crystal limit the phase-matching function
couples $\mathbf Q$ and $\mathbf q_1-\mathbf q_2$ and the factorization fails; the
idler is then the cross-modulated field blurred by the phase-matching
point-spread function $\tilde\gamma$, as made explicit in
Eq.~\eqref{eq:pump_transfer_conv}. Equation~\eqref{eq:product} therefore requires
that the crystal be short compared with the Rayleigh ranges of the interacting
beams, which in practice means that the transverse structure to be transferred
must be coarse compared with the phase-matching resolution.
We note that the product relation fixes the idler field \emph{at the
crystal plane}. The idler subsequently propagates with wavenumber $k_i$, not
$k_p$ or $k_s$. Since Fresnel propagation depends on the ratio $k/z$, a structure
that the pump carried a distance $z_p$ before the crystal reappears in the idler
at $z_i = (k_i/k_p)\, z_p$, and a wavefront that the seed acquired over $z_s$ is
undone by the conjugate idler after $z_i = (k_i/k_s)\, z_s$. Image transfer and
phase conjugation in StimPDC are thus always accompanied by a longitudinal
rescaling set by the wavelength ratio, a point first made explicit
experimentally in Ref.~\cite{souto2001observation}.

The remainder of this review is, to a large extent, an exploration of
Eq.~\eqref{eq:product} in successively richer settings, summarized in
Table~\ref{tab:product}.

\begingroup
\squeezetable
\begin{table}[t]
\caption{Consequences of the product relation, Eq.~\eqref{eq:product}.}
\label{tab:product}
\begin{ruledtabular}
\begin{tabular}{lll}
Limit or generalization & Result & Section \\
\hline
$\mathcal E_s\rightarrow$ const & $\mathcal A_i \propto \mathcal E_p$ &
\ref{subsec:asit} \\
$\mathcal E_p\rightarrow$ const & $\mathcal A_i \propto \mathcal E_s^{*}$ &
\ref{subsec:asit} \\
$LG$ inputs, azimuthal part & $\ell_i=\ell_p-\ell_s$ &
\ref{subsec:OAMconservation} \\
$LG$ inputs, radial part & multi-$p$ superposition &
\ref{sec:oam} \\
vortex $\times$ triangular beam & triangular lattice &
\ref{sec:oam} \\
two-crystal, vector inputs & vector phase conjugation &
\ref{sec:vvb} \\
seed replaced by AWP field & advanced-wave picture &
\ref{subsec:awp} \\
partial coherence & $W_i = W_p W_s^{*}$, twist conservation &
\ref{subsec:tgsm} \\
\end{tabular}
\end{ruledtabular}
\end{table}
\endgroup

\subsubsection{Amplitude transfer from the seed: Phase Conjugation}
\begin{figure}[t]
\centering
\includegraphics[width=\linewidth]{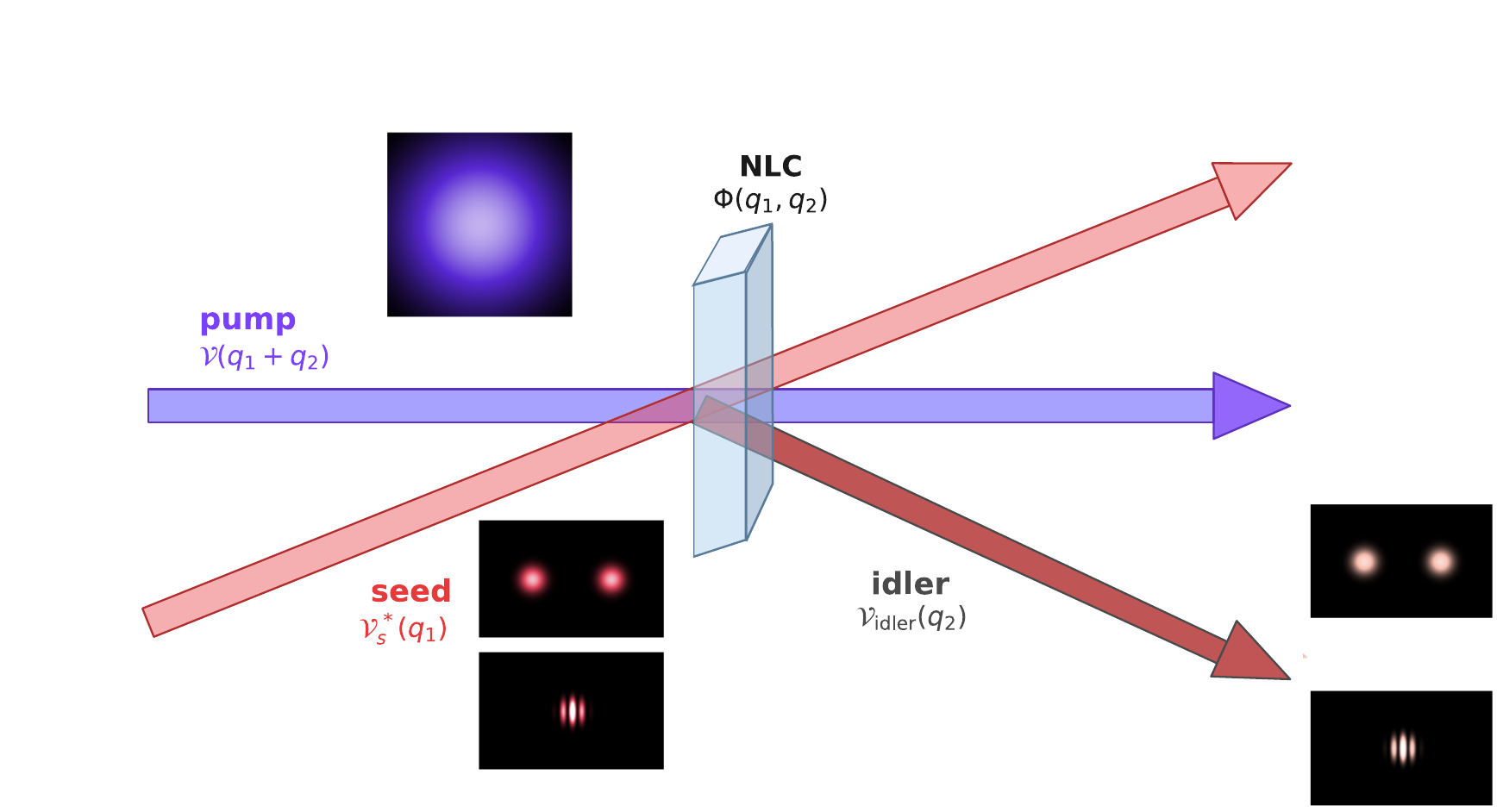}
\caption{Transfer of seed angular spectrum $v_s^*(\mathbf q_2)$ via StimPDC. The insets illustrate pump intensity, seed and idler intensity (upper patterns) and corresponding Fourier Transforms (lower patterns) for seed and idler.}
\label{fig:seed_transfer}
\end{figure}

From Eq.~\eqref{eq:product} we demonstrate explicitly that, in the thin-crystal limit, the field amplitude of the idler is the product of the pump field with the complex conjugate of the seed field. Let us further apply a plane wave approximation for the pump field:
$v(\mathbf q_1+\mathbf q_2) \propto \delta(\mathbf q_1+\mathbf q_2)$, or equivalently $\mathcal E_p(\boldsymbol\rho)=\mathrm{constant}$.
In this case the idler field is given by $\mathcal A_{\mathrm{stim}}(\boldsymbol\rho)
\;\propto\;\mathcal E_s^{*}(\boldsymbol\rho)$, or in terms of angular spectrum:
\begin{equation}
v_\text{idler}(\mathbf q_2) \propto v_s^*(-\mathbf q_2).
\label{eq:idler_amplitude_conj-seed}
\end{equation}
The idler field is the phase conjugate of the seed field, which is a consequence of the spatial entanglement properties of the source in the SPDC regime. Consequently, any spatial structure or wavefront distortion encoded in the seed is faithfully transferred to the idler with an inverted phase profile, which leads to several applications. In particular, there is the intriguing correction of wavefront distortion demonstrated with phase-conjugation mirrors in the
1970s~\cite{zeldovich72,boyd_nlo} and, in the four-wave-mixing platform, with
entangled and phase-conjugate image pairs from atomic
vapors~\cite{Boyer2008Images}, and more recently in StimPDC~\cite{Cardoso26}. Moreover, the seed and idler can have quite different wavelengths, making the idler a complex conjugated replica of the seed at a different optical frequency. The transfer of image and angular spectrum from the seed to the idler is illustrated in Fig.~\ref{fig:seed_transfer}. 

\subsubsection{Amplitude transfer of Pump-Phase-Matching}
Let us return to Eq.~\eqref{eq:idler_field}, where the idler field depends upon the angular spectrum of the seed and pump, as well as the phase matching function.
Making the plane wave approximation this time for the seed field, and recalling that the Dirac delta function is real and symmetric so that its complex conjugate remains $v_s^*(\mathbf q_1) \propto \delta(\mathbf q_1)$, or equivalently $\mathcal{E}^*_s= \mathrm{constant}$, the idler becomes
\begin{equation}
v_\text{idler}(\mathbf q_2) \propto v(\mathbf q_2) \gamma(-\mathbf q_2).
\end{equation}
To describe then the idler in position space, we use the convolution theorem: the inverse Fourier transform of a product of two functions is equivalent to the convolution of their individual inverse Fourier transforms. Let $\mathcal{E}_p(\boldsymbol{\rho})$ denote the real-space spatial amplitude of the pump beam, and $\tilde{\gamma}(\boldsymbol{\rho})$ denote the inverse Fourier transform of the phase-matching function $\gamma$. The real-space idler field amplitude can thus be expressed as:
\begin{equation}
\mathcal{A}_{\text{idler}}(\boldsymbol{\rho}_2) \propto \mathcal{E}_p(\boldsymbol{\rho}_2) * \tilde{\gamma}(\boldsymbol{\rho}_2),
\label{eq:pump_transfer_conv}
\end{equation}
where $*$ represents the 2-D spatial convolution operator. 

\begin{figure}[t]
\centering
\includegraphics[width=\linewidth]{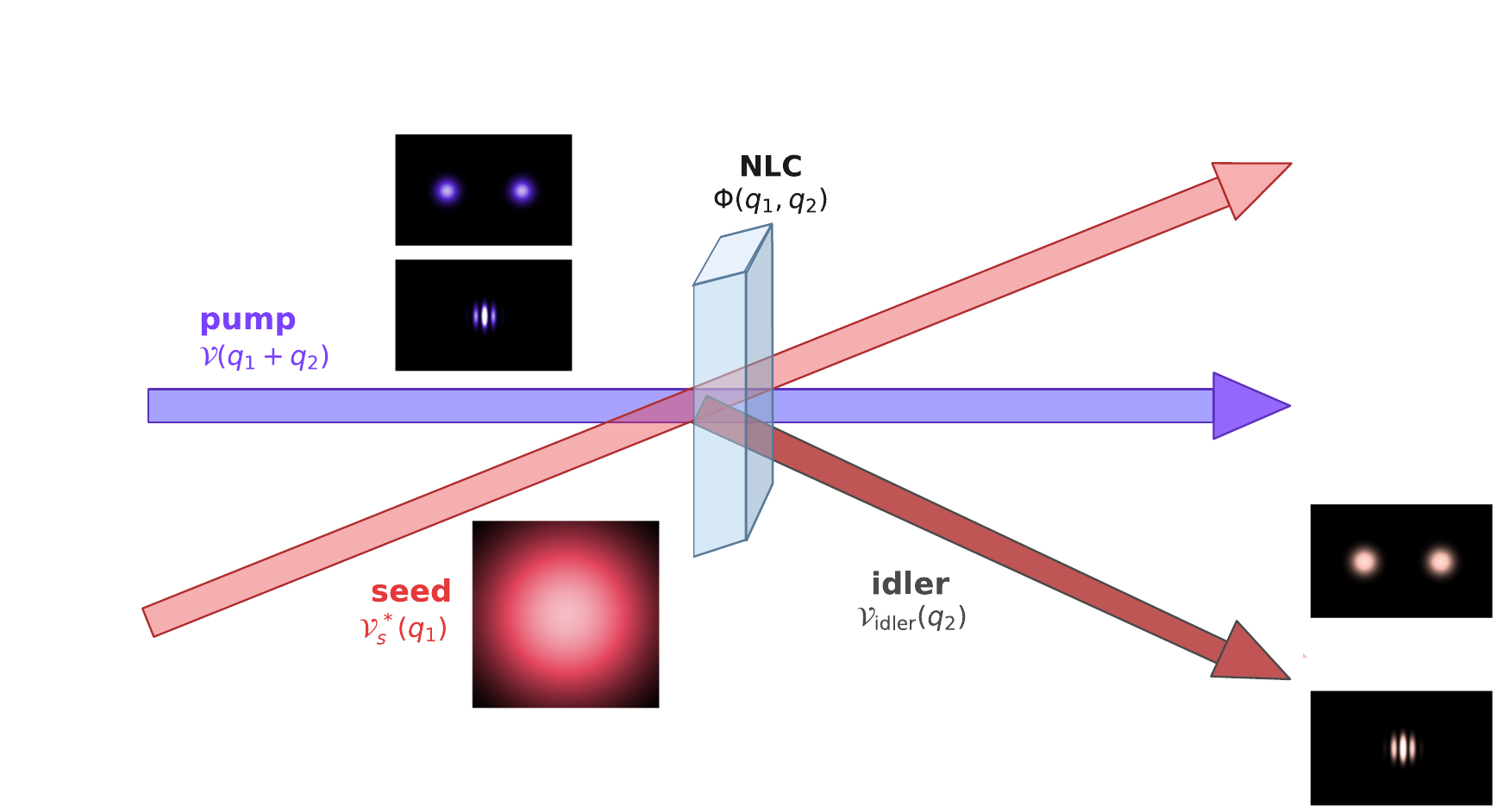}
\caption{Transfer of pump-phase-matching angular spectrum $\Phi(\mathbf q_1,\mathbf q_2) = v(\mathbf q_1+\mathbf q_2)\,\gamma(\mathbf q_1- \mathbf q_2)$ via StimPDC. The insets illustrate seed intensity, pump and idler intensities (upper patterns) and corresponding Fourier Transforms (lower patterns).}
\label{fig:pump_transfer}
\end{figure}

Equation~\eqref{eq:pump_transfer_conv} dictates that the idler beam forms a spatial image of the pump beam, blurred by the crystal's phase-matching point-spread function $\tilde{\gamma}(\boldsymbol{\rho}_2)$. Under the thin-crystal approximation, the phase-matching bandwidth becomes infinitely broad ($\gamma \rightarrow 1$), and its inverse Fourier transform approaches a Dirac delta function, $\tilde{\gamma}(\boldsymbol{\rho}_2) \propto \delta(\boldsymbol{\rho}_2)$. In this limit, the convolution resolves to:
\begin{equation}
\mathcal{A}_{\text{idler}}(\boldsymbol{\rho}_2) \propto \mathcal{E}_p(\boldsymbol{\rho}_2),
\label{eq:idler_amplitude}
\end{equation}
demonstrating an exact, aberration-free real-space image transfer from the pump to the idler beam. The idler thus carries a wavelength-converted replica of the pump angular
spectrum. See Fig.~\ref{fig:pump_transfer} illustrating the process.

\subsection{Coherence transfer}
\label{subsec:coherence}

The coherence properties of the idler beam are defined by the product between pump and seed amplitudes and it is also affected by the incoherent contribution coming from spontaneous emission. In the simplest case, where pump and seed are perfectly coherent, the degree of coherence of the idler can be derived in the following way, as described in Ref.~\cite{Ribeiro95}.

The light produced in StimPDC is partially coherent in general, because of the mixing between spontaneous and stimulated emissions. In order to calculate the degree of coherence, we suppose that the electric field at position $r$ has two contributions: the coherent field component $E_c$ and an incoherent field component $E_i$:
\begin{equation}
E(r) = E_c(r) + E_i(r).
\end{equation}
The spatial coherence of the beam is quantified by the modulus of the normalized mutual intensity, $|\mu_{12}|$, which directly determines the visibility of interference fringes in a Young's double-slit experiment:
\begin{equation}
\mu_{12} = \frac{\langle E^*(r_1) \cdot E(r_2) \rangle}{\sqrt{\langle E^*(r_1) \cdot E(r_1) \rangle \langle E^*(r_2) \cdot E(r_2) \rangle}}.
\end{equation}

Because the phases of the purely incoherent component fluctuate randomly while the stimulated coherent field maintains a stable phase, cross-correlation terms between $E_c$ and $E_i$ average to zero. Consequently, the coherence function simplifies to a weighted sum of the incoherent and coherent contributions:
\begin{equation}
\mu_{12} = \frac{\mu_i I_i + \mu_c I_c}{I_i + I_c},
\end{equation}
where $\mu_i$ and $\mu_c$ are the coherence functions of the incoherent and coherent fields, while $I_i$ and $I_c$ represent their respective intensities. Notice that the incoherent field at the source gives rise to partially coherent light after propagation according to the Van Cittert--Zernike theorem.

By introducing the average occupation number per mode, $\mathcal{N}$, defined as the ratio of coherent to incoherent intensities ($\mathcal{N} = I_c / I_i$), the expression for the normalized mutual intensity can be rewritten as a function of the stimulation strength:
\begin{equation}
\mu_{12} = \frac{\mu_i + \mu_c \mathcal{N}}{\mathcal{N} + 1}.
\end{equation}

The average occupation number $\mathcal{N}$ is directly proportional to the photon rate (intensity) of the seed laser through the coupling parameter $\beta$:
\begin{equation}
\mathcal{N} = \beta \langle n \rangle,
\end{equation}
where $\langle n \rangle$ is the mean photon number of the seed laser within one coherence volume. 

As the intensity of the stimulation laser increases ($\langle n \rangle \to \infty$, hence $\mathcal{N} \to \infty$), the mutual intensity approaches that of the coherent component ($\mu_{12} \to \mu_c$). This demonstrates that tuning the intensity of the seed laser directly controls the spatial coherence of the down-converted beam, causing it to transition from a thermal-like (incoherent) statistical regime to a laser-like (spatially and temporally coherent) regime.

\section{Spatial Coherence, Image and angular spectrum Transfer}
\label{sec:fundamentals}

We review some pioneering and foundational works that established how spatial coherence properties, interference patterns, and images are transferred from either the pump beam or the seeding laser beam to the stimulated idler field. They either experimentally demonstrated or theoretically described the correlations explained in Sec.~\ref{sec:theory}, launching important developments that will be outlined in the remainder of this paper.

The ability to transfer spatial information and coherence properties from a pump or seed field to the generated down-converted fields is a defining characteristic of StimPDC. Unlike the spontaneous regime, where the spatial correlations are fundamentally governed by vacuum fluctuations and phase-matching conditions, the stimulated regime allows for the deterministic mapping of complex wavefronts, facilitating the observation of the correlations.

One of the seminal works demonstrating the presence of correlations in the stimulated regime similar to what have been previously observed in the spontaneous regime is Ref.~\cite{Ribeiro95}. By pumping a $\text{LiIO}_3$ crystal with a 351.1~nm laser and seeding the nonlinear conversion with a 632.8~nm laser of variable power, the authors controlled the fringe visibility of the double-slit interference pattern formed at the idler beam. They observed an improvement of the visibility with seed power, and associated the results with the occupation number per mode computed from the ratio of coherent (stimulated) and incoherent (spontaneous) intensities. The occupation number increased linearly with seed power, while the transverse coherence (evaluated by the fringe visibility) showed a transition from the spontaneous to stimulated emission, in accordance with the behavior expected from the occupation number per mode and the mutual intensities (between slits) for the coherent and incoherent fields. Such nonlocal control of the visibility had already been reported when a double slit was placed at the signal beam in the spontaneous regime~\cite{ribeiro1994}.

A formal quantum-mechanical description of the previous results was then introduced in Ref.~\cite{ribeiro1999}, whose underlying concepts and consequences, such as the direct image transfer from the pump to the stimulated field and the conjugate image transfer from the seed to the stimulated field, are mostly covered in Sec.~\ref{sec:theory}. In addition, it provided a fundamental explanation of the characteristic relation of the visibilities observed in Ref.~\cite{Ribeiro95} and explained the coherence transfer from the seed to the stimulated field (adding up to the incoherent background stemming from spontaneous emission). In particular, the double-slit interference pattern of the idler field from line-like sources as pump and seed beams was computed, and the authors showed that the visibility is given by the Van Cittert--Zernike theorem for spatially incoherent light.

Building on such theoretical unfolding, the authors of Ref.~\cite{souto2001observation} observed the expected image transfer and phase conjugation in StimPDC. They compared the images, recorded downstream from the crystal plane, of the idler beam and either pump or seed fields modulated by double slits. With that, they demonstrated the direct image transfer from pump to stimulated (idler) field, but with the propagation vector defined by the idler wavelength, as given by Eq.~\eqref{eq:idler_amplitude}. By further utilizing a lens between the double slits and the crystal, they
accessed the conjugate relation in the Fourier domain, where the reversal
$\mathbf q_2 \rightarrow -\mathbf q_1$ implied by
Eq.~\eqref{eq:idler_amplitude_conj-seed} appears as an inversion of the idler
distribution with respect to the seed. At the crystal plane itself the
conjugation leaves the intensity unchanged and acts only on the phase. It is worth noting that the type-II, non-degenerate geometry has the additional advantage that signal and idler differ both in polarization and
in wavelength, so scattered or transmitted seed light can be rejected independently
by a polarizer and by a spectral filter.

A more recent contribution on the investigation of spatial coherence in StimPDC is given in Ref.~\cite{santos2023}. In this work, the authors theoretically introduce a connection between the mutual coherence of the idler field with (i) the weighted contributions from spontaneous and stimulated emission~\cite{Ribeiro95}, and (ii) the mutual coherence of spontaneous emission described by the cross-spectral density of partially coherent light fields of the Gaussian-Schell-Model type~\cite{Hutter20,Hutter21}. The experimental challenge of measuring the visibility as a function of separation distance between points within the wavefront was overcome by utilizing a misaligned Michelson interferometer for the idler beam, so that the separation was controlled by controlling the mirror tilt angle of the interferometer. The resulting visibility was further measured as a function of seed power, and, by taking a threshold value for it, the authors obtained the observed transverse coherence length in the transition from the spontaneous (partial coherence) to the stimulated regime (increasingly coherent).

\section{Structured Light and OAM}
\label{sec:oam}

Image transfer, coherence transfer, and phase conjugation through three-wave mixing were extensively investigated in the late 1990s and early 2000s~\cite{souto2001observation}. While the conservation of orbital angular momentum (OAM) in spontaneous parametric down-conversion~\cite{arlt1999parametric}, optical pumping of cold atoms~\cite{tabosa1999optical}, and frequency up-conversion~\cite{dholakia1996second} had already been established, the behavior of structured beams in the stimulated regime remained unexplored until 2002. In particular, fundamental questions concerned the conservation of topological charge, the transfer of complex spatial phases, and the influence of nonlinear interactions on the spatial-mode structure of the generated fields.

This section is organized around three questions that were opened in sequence:
(1) whether topological charge is conserved and conjugated in the stimulated regime, (2)
whether conservation of the azimuthal index implies conservation of the full
spatial mode and (3) what happens when neither input beam is a pure Laguerre--Gauss
mode. In short, the answers are, respectively, yes, no, and a rich family of
cross-modulation patterns, and below we show why. These studies reveal how three-wave mixing governs the coupling of structured optical fields, leading to phenomena such as topological-charge inversion, the generation of new radial modes, and the emergence of geometric intensity patterns, including triangular lattices. Together, they establish a comprehensive picture of spatial-mode dynamics in stimulated nonlinear optical interactions.

Light beams carrying orbital angular momentum can be described by Laguerre--Gauss modes, denoted by $LG_m^\ell$, where $m$ is the radial number and $\ell$ the topological charge. These are cylindrical solutions of the paraxial wave equation characterized by an azimuthal phase $\exp(i\ell\theta)$, where each photon in such a mode carries an orbital angular momentum $\ell\hbar$~\cite{Allen1992,padgett2000light}. At the crystal plane, the radial envelope of these modes is real, so complex conjugation of the field acts only on the azimuthal phase: a mode of charge $\ell$ conjugates into a mode of charge $-\ell$. This fact, together with the phase-conjugation relation already established in Sec.~\ref{sec:theory} underlies every result discussed below. 

\subsection{Conservation and phase conjugation of the topological charge}
\label{subsec:OAMconservation}

From the amplitude product relation~\eqref{eq:product}, in the thin crystal limit the idler amplitude generated inside the nonlinear crystal is proportional to the product of the pump and the complex conjugate of the seed.
When the pump and the seed beams are prepared in Laguerre--Gauss modes carrying topological charges $\ell_{p}$ and $\ell_{s}$, the idler amplitude has an azimuthal dependence given by $e^{i(\ell_{p}-\ell_{s})\phi}$, so the generated idler must carry the topological charge
\begin{equation}\label{eq:l_conservation}
    \ell_{i} = \ell_{p}-\ell_{s}.
\end{equation}

The conservation of topological charge follows from the rotational invariance of the three-wave mixing Hamiltonian~\cite{dosSantos2025TwistConservation}. Since the nonlinear interaction cannot create or destroy orbital angular momentum in a rotationally symmetric medium, the total OAM must be conserved during the conversion process. This symmetry is valid under the same paraxial and near-collinear regime assumed throughout Sec.~\ref{subsec:idlerStimPDC}. 

\begin{figure}[t]
\centering
\includegraphics[width=\linewidth]{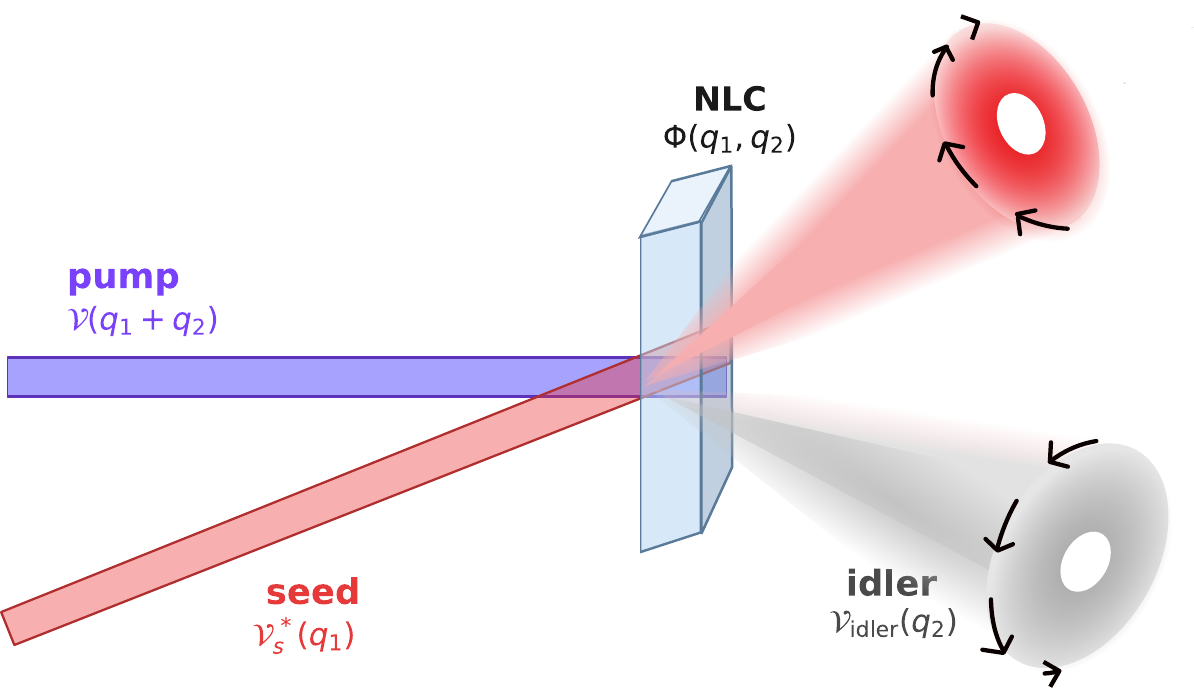}
\caption{Illustration of stimulated emission driven by structured light. A Laguerre--Gauss mode is used to stimulate emission in a nonlinear crystal pumped with a Gaussian blue laser light. The generated idler beam acquires the same topological charge as the seed with opposite sign.}
\label{fig:OAM_stim}
\end{figure}

The first experimental verification of this relation was developed in~\cite{caetano2002conservation}. Using a type-II BBO crystal pumped at $442$~nm and a stimulating beam prepared in a $LG_{0}^{1}$ mode via computer-generated hologram, they measured the idler's topological charge with a misaligned Michelson interferometer. This technique was necessary because with only the beam intensity, it is not possible to distinguish between $-\ell$ and $+\ell$. With this setup, the idler's topological charge is inferred from the fork dislocation. By preparing a pump with $\ell_p=+1$ and a seed with $\ell_s=0$, the inferred idler topological charge is $\ell_i=+1$; when a seed makes the preparation with $\ell_{s}=+1$ and a pump $\ell_p=0$, the resulting conjugation is confirmed with $\ell_i = -1$. Figure~\ref{fig:OAM_stim} illustrates this result. This experiment provided the first direct demonstration that the OAM conservation law predicted for spontaneous PDC also remains valid in the stimulated regime, where the generated idler can be directly characterized.

A complementary, geometric picture of the same conjugation was later developed by de Oliveira et al.~\cite{de2019phase}, who generalized the discrete charge flip to continuous superpositions of first-order modes. Writing the seed as $E_s(\rho) = \cos(\theta/2)LG_{0}^{+1}(\rho)+e^{i\phi}\sin(\theta/2)LG_{0}^{-1}(\rho)$, each pair ($\theta,\phi$) labels a point on a Poincaré sphere for spatial modes, where the poles correspond to pure $LG_{0}^{\pm 1}$ vortices and the equator to Hermite--Gauss modes, which are superpositions of LG modes. The polar angle controls the relative weights of the two OAM eigenstates, whereas the azimuth determines their relative phase. 

Applying the conjugation rule above to this superposition, the idler is found at $(\theta_{i}, \phi_{i}) = (\pi-\theta, \phi)$: the mirror reflection of the seed's point through the equatorial plane. The special case $\theta= 0$ recovers $\ell_{i} = -\ell_s$, connecting this geometric picture to the discrete result of Caetano et al.~\cite{caetano2002conservation}. The trajectory was verified experimentally with a tilted-lens method, and the
same authors related the reflection both to Klyshko's advanced-wave picture
(Sec.~\ref{subsec:awp}) and to the mode conversion occurring in optical
parametric oscillators, where cavity boundary conditions impose additional
constraints~\cite{rodrigues2018orbital}.

While the azimuthal index follows a simple conservation law, the radial degree of freedom behaves differently and generally couples to multiple modes, motivating the discussion in the next section.

\subsection{Beyond conservation: radial-mode generation}

The conservation law of Eq.~\eqref{eq:l_conservation} constrains only the azimuthal degree of freedom of the generated field. While it uniquely determines the idler topological charge, it does not specify how the radial profile evolves during the nonlinear interaction. This question was addressed by de Oliveira \textit{et al.}~\cite{Oliveira21}, who investigated whether StimPDC preserves the complete Laguerre--Gauss (LG) mode structure or whether the nonlinear interaction can populate additional radial modes.

To isolate this effect, both the pump and the stimulating beam were prepared in pure Laguerre--Gauss modes with null radial order, $LG_m^{\ell}$ with $m=0$. Under the amplitude product relation in the thin-crystal limit~\eqref{eq:product}, the idler field is proportional to the product of pump and seed amplitudes, 
so that its radial dependence scales as $r^{|\ell_p|+|\ell_s|}$. Consequently, conservation of the azimuthal index alone is generally insufficient to preserve the complete LG modal structure.

A pure Laguerre--Gauss mode is recovered only when
\begin{equation}
|\ell_i|=|\ell_p|+|\ell_s|,
\end{equation}
which occurs for counter-rotating vortices ($\ell_p\ell_s<0$). In this situation, the idler remains a single $LG_0^{\ell_i}$ mode and propagates self-similarly. Conversely, for co-rotating vortices ($\ell_p\ell_s>0$), one finds
\begin{equation}
|\ell_i|<|\ell_p|+|\ell_s|,
\end{equation}
so that the radial dependence of the generated field is incompatible with that of a single LG mode. The idler must therefore be expanded as a coherent superposition
\begin{equation}
LG_p^{\ell_i},
\qquad
p=0,\ldots,P,
\end{equation}
with
\begin{equation}
P=\min\!\left(|\ell_p|,|\ell_s|\right),
\end{equation}
revealing that the nonlinear interaction generates radial-mode content even though both input beams possess trivial radial order.

The physical origin of this effect is that the nonlinear interaction preserves only the total azimuthal phase imposed by rotational symmetry, while the radial dependence arises from the overlap of the transverse profiles of the interacting fields. Therefore, unlike the topological charge, the radial index is not protected by a conservation law and can be redistributed among several LG modes without violating orbital angular momentum conservation.

Since each radial component accumulates a different Gouy phase,
\begin{equation}
-(2p+|\ell_i|+1)\arctan(z/z_R),
\end{equation}
their relative phases evolve during propagation. As a result, the near-field intensity remains ring-shaped for both co- and counter-rotating vortices. In contrast, the radial-mode decomposition becomes evident only after free-space propagation toward the far field, where interference between the different Gouy phases produces multiple concentric rings.

The theoretical predictions were experimentally verified using a type-I BBO crystal, with both the pump and seed beams prepared by spatial light modulators under conditions that satisfied the thin-crystal approximation. The measurements confirmed the conservation of the azimuthal charge for every combination of input modes while revealing multiple radial rings exclusively for co-rotating vortices, in excellent agreement with the theoretical model. Interestingly, this behavior is opposite to that observed in up-conversion generation, where radial-mode generation occurs for counter-rotating vortices~\cite{Pereira17,buono2020chiral}. This inversion originates from the phase conjugation inherent to the StimPDC interaction, which reverses the role played by the azimuthal phase in the nonlinear mode conversion.

This result represents an important extension of the OAM conservation law. It demonstrates that conserving the topological charge does not imply conservation of the complete spatial mode and establishes StimPDC as a mechanism capable of coherently redistributing optical power among radial Laguerre--Gauss modes. As discussed in the next section, this discrete mode-conversion picture is itself a particular manifestation of a more general interpretation, in which the idler field can be understood as arising from the diffraction of one structured beam through the transverse profile of the other.
The corresponding statement in the high-gain regime is that the correlations
between twin beams carrying orbital angular momentum become delocalized over the
transverse profile~\cite{Marino2008OAM}, discussed briefly in Sec.~\ref{subsec:highgain}.

\subsection{Structured light as spatial cross-modulation: the triangular lattice}

The two preceding subsections treated inputs that were pure Laguerre--Gauss
modes, for which the amplitude product relation in StimPDC~\eqref{eq:product} reduces to a statement about indices. When
one of the two beams is not a vortex at all, Eq.~\eqref{eq:product} is better read
literally, as the modulation of one transverse profile by the other. In that
reading StimPDC becomes an all-optical spatial modulator whose transfer function
is written by the second beam.

Pinheiro da Silva \emph{et al.}~\cite{Santos2022} realized this idea with a
vortex beam of charge $\ell$ and a triangular-shaped beam. The far field of the
converted beam is a triangular lattice of intensity maxima, the optical analog
of the well-known diffraction of a vortex by a triangular aperture~\cite{Hickmann2010,Hickmann_2011}, but produced
here by nonlinear cross modulation rather than by a physical mask. Both the
number of lattice sites and the orientation of the lattice are set by the
topological charge of the vortex, so the pattern constitutes a direct,
single-shot readout of $\ell$, including its sign. The authors examined both
up-conversion and down-conversion. In the down-converted case the phase
conjugation carried by Eq.~\eqref{eq:product} enters explicitly: because the seed
contributes as $\mathcal E_s^{*}$, reversing which of the two beams carries the
vortex reverses the effective sign of the charge, and the lattice orientation
rotates accordingly. 

\begin{figure}[t]
\centering
\includegraphics[width=\linewidth]{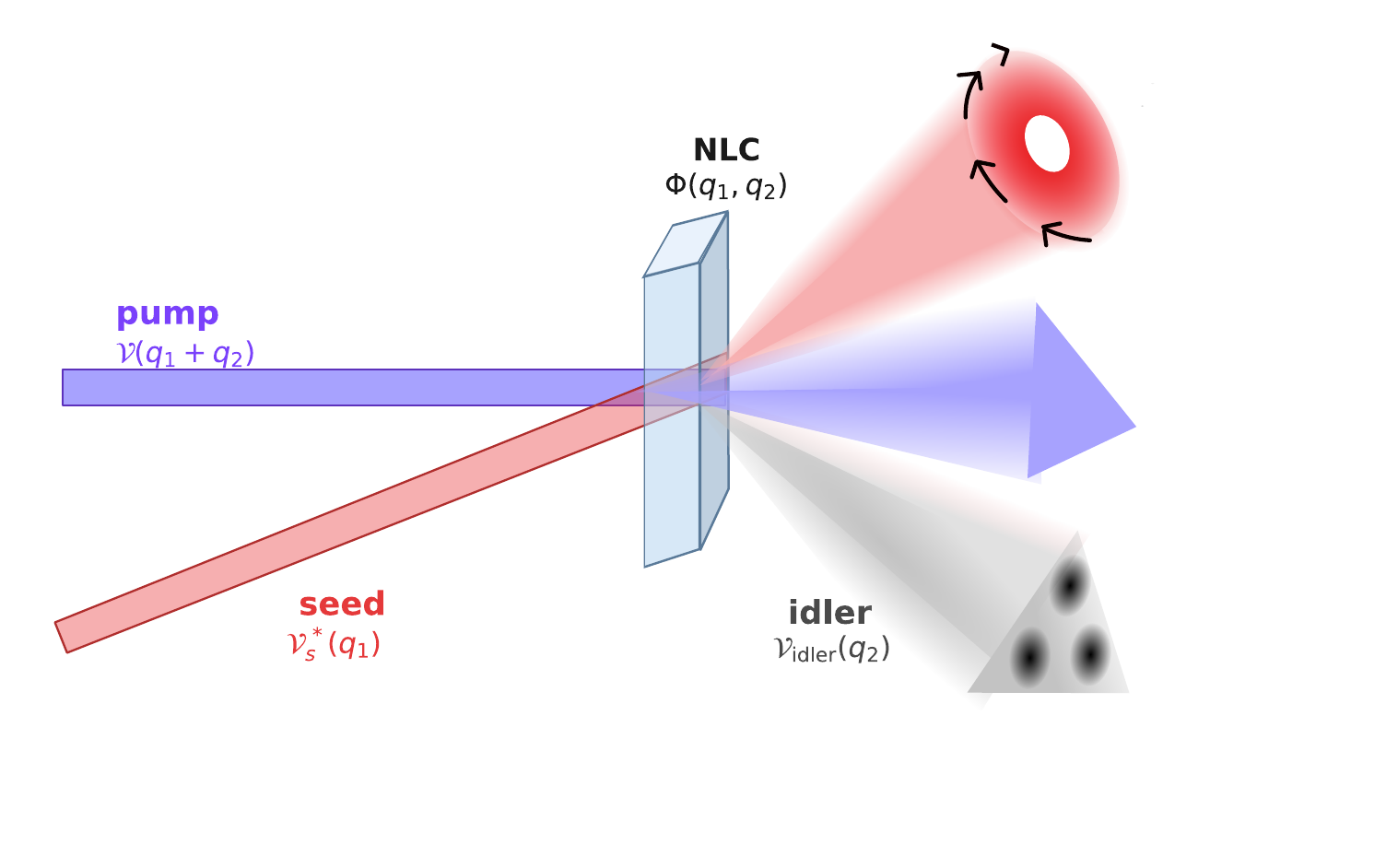}
\caption{Cross-modulation of light fields in stimulated down-conversion. A triangular-shaped beam pumps the nonlinear medium, while a Laguerre--Gauss beam stimulates the down-conversion process. The resulting idler beam carries the product of the two beams.}
\label{fig:trian_CM}
\end{figure}

Two features of this experiment matter for what follows. The first is
generality: the cross modulation is not tied to vortex--triangle mixing, and
works for a wide variety of transverse structures, so the same mechanism can
implement in a crystal what would otherwise require a spatial light modulator at
a wavelength where one might not exist. The second is a connection to wavefront
filtering. Rocha \emph{et al.}~\cite{Rocha2021Speckle} showed that mixing noisy
modes in a $\chi^{(2)}$ medium can filter a random phase without
correlation-based post-processing, but their sum-frequency configuration requires
the conjugate phase to be present in one of the two input patterns. In StimPDC
that restriction is lifted automatically, since the conjugation is supplied by
the interaction itself. This suggests using a random phase screen as an
encryption key: information encoded on the seed is transferred to the idler, and
is recovered by propagating the idler back through the same screen. We return to
this in Sec.~\ref{sec:outlook}.

\section{Vector Vortex Beams}
\label{sec:vvb}

The previous sections have shown that stimulated parametric down-conversion
(StimPDC) faithfully reproduces spatial information, enabling image transfer,
mode conversion and orbital angular momentum (OAM) processing through the
nonlinear interaction. A natural question that follows is whether these
capabilities can be extended beyond scalar structured light. Can StimPDC
process \emph{vector} beams, reproducing not only their amplitude and phase,
but also their spatially varying polarization structure? 

Vector vortex beams (VVBs) provide an ideal platform to address this question.
Unlike scalar vortex beams, whose polarization is homogeneous across the
transverse plane, VVBs combine orbital angular momentum with spatially varying
polarization, resulting in intrinsically non-separable spin--orbit
states~\cite{Borges10,Holleczek11}.
Radially and azimuthally polarized beams constitute the simplest examples, while more general hybrid vector beams exhibit spatial distributions containing linear,
elliptical and circular polarization states simultaneously~\cite{Milione2011,Zhan09}. This hybrid structure has established VVBs as
classical analogues of Bell states~\cite{Borges10,Holleczek11} and as
versatile resources for spin--orbit engineering, high-dimensional optical
encoding, precision metrology and optical communications~\cite{Souza2008,DAmbrosio2012,DAmbrosio2013,Zhao2015}.

The extension of StimPDC to this class of structured fields is not trivial.
Conventional type-I nonlinear interactions couple only one polarization
component, preventing the coherent conversion of arbitrary vector beams.
This limitation was overcome through the use of the two-type-I-crystal
sandwich configuration, originally developed as a high-brightness source of polarization-entangled photon pairs~\cite{Kwiat1999}. In this geometry, two orthogonally oriented nonlinear crystals independently mediate the interaction for
horizontal and vertical polarization components while preserving coherence
between them.

\begin{figure}[t]
\centering
\includegraphics[width=\columnwidth]{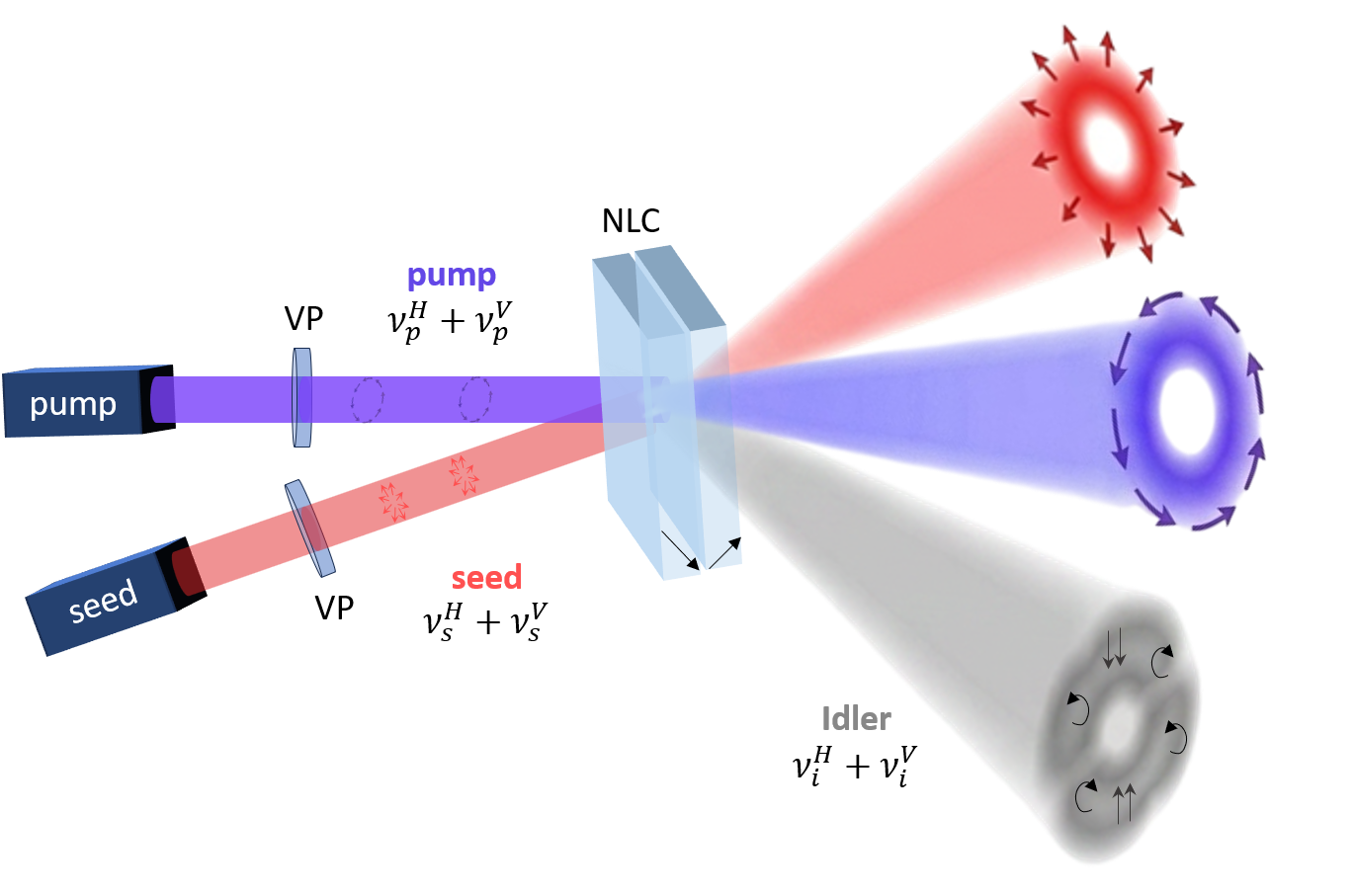} 
\caption{Experimental scheme for realizing two-crystal StimPDC with interacting vector vortex beams.}
\label{Vec}
\end{figure}

This architecture enabled real-time phase conjugation of
vector vortex beams via StimPDC, demonstrating that the nonlinear process is
capable of simultaneously transferring amplitude, phase and polarization
information~\cite{deoliveira2020realtime}. A complete quantum-optical description
was subsequently developed, revealing that the polarization transformation of
the stimulated idler is directly determined by the polarization-entangled
state that the same source would generate in spontaneous parametric
down-conversion~\cite{deoliveira2020quantum}. Finally, a comprehensive
experimental characterization confirmed these predictions through transverse
polarization tomography, establishing StimPDC as a reliable platform for
manipulating hybrid spin--orbit optical fields~\cite{daSilva2021PRA}.

\subsection{StimPDC with vector vortex beams}

Within the paraxial approximation, cylindrical vector beams may be described
as coherent superpositions of counter-rotating circularly polarized
Laguerre--Gaussian modes~\cite{Milione2011,deoliveira2020quantum}. Because the
two-crystal configuration supports nonlinear interactions for both
polarization components simultaneously, every component of the pump and seed
fields contributes coherently to the generated idler beam.

Consequently, StimPDC no longer performs only spatial mode transfer but rather
implements a nonlinear transformation acting on the complete spin--orbit
structure of the optical field. Depending on the pump and seed
configurations, the generated idler may preserve radial or azimuthal
polarization, generate hybrid vector states or evolve into more complex
polarization textures arising from coherent superpositions of paraxial modes
with different Gouy phases~\cite{daSilva2021PRA}.

\subsection{Vector phase conjugation}
\label{subsec:vpconj}

One interesting consequence of the two-crystal configuration is
that StimPDC naturally implements vector phase conjugation. While scalar phase
conjugation reverses only the transverse phase profile, vector phase
conjugation simultaneously operates on spatial and polarization degrees of
freedom.

The quantum-optical framework developed in Ref.~\cite{deoliveira2020quantum} demonstrated that the stimulated process is equivalent to a projection onto the polarization state of the seed beam, establishing a direct correspondence between vector phase conjugation and the Bell-state transformations produced by the same nonlinear source.

Within the Stokes representation, conventional vector phase conjugation by an entangled-photon source producing the $(\ket{HH}+\ket{VV})/\sqrt{2}$ state (in the SPDC regime) is described by
\begin{equation}
(S_1,S_2,S_3)\rightarrow(S_1,S_2,-S_3),
\end{equation}
corresponding to a mirror reflection across the equatorial plane of the
Poincaré sphere. More generally, different Bell-state configurations produce
conjugation operations defined with respect to different polarization bases,
while one particular configuration realizes a universal NOT operation (see Sec.~\ref{sec:quantum}), providing an elegant connection between nonlinear optics and quantum information processing~\cite{deoliveira2020quantum}.

\subsection{Experimental characterization}

The theoretical predictions were experimentally verified through complete
transverse polarization tomography of the stimulated idler
beam~\cite{daSilva2021PRA}. By recording projections onto the
horizontal--vertical, diagonal--antidiagonal and right--left circular
polarization bases, the local Stokes parameters were reconstructed over the
entire transverse profile.

The measurements showed good agreement with the theoretical model for a
wide variety of pump and seed configurations. In particular, radially and
azimuthally polarized beams preserve their vectorial character after the
nonlinear interaction, while hybrid vector beams generate intricate spatial
polarization distributions that evolve during propagation due to coherent
superpositions of modes with different Gouy phases~\cite{daSilva2021PRA}.

These observations also have implications beyond the stimulated regime.
Because spontaneous and stimulated parametric down-conversion share identical
phase-matching conditions and mode-coupling mechanisms, StimPDC provides a
powerful classical platform for investigating and visualizing phenomena in multiple degrees of freedom that would otherwise require single-photon measurement devices and detectors in SPDC.

\subsection{Outlook}
\label{subsec:vvboutlook}

The incorporation of vector vortex beams significantly expands the scope of
StimPDC. Rather than acting exclusively as a mechanism for image transfer or
orbital angular momentum conversion, the nonlinear interaction becomes a
general platform for processing structured light in its complete vectorial
form.

The simultaneous manipulation of amplitude, phase and polarization enables
frequency conversion while preserving spin--orbit structure, opening
perspectives for mode-selective amplification, adaptive optics, structured
light engineering, high-dimensional communications and analog simulations of
hybrid quantum systems. Together with the developments discussed in the
following sections on nonlinear imaging, holographic modeling and partially
coherent beams, these results establish StimPDC as a versatile framework for
the nonlinear processing of increasingly complex optical fields.

\section{Classical Characterization, Nonlinear Imaging, and Partial Coherence in Stimulated Down-Conversion}
\label{sec:nonlinear}

The preceding sections built up the theoretical framework of StimPDC [Sec.~\ref{sec:theory}] and used it to describe how the transverse structure of the pump and seed is coherently mapped onto the stimulated idler---first for scalar structured light [Secs.~\ref{sec:fundamentals}--\ref{sec:oam}] and then for vector fields carrying spatially varying polarization [Sec.~\ref{sec:vvb}]. In each of those cases, StimPDC was studied as a process whose transfer laws were the object of investigation. We now shift perspective and consider StimPDC as a tool: a bright, mode-selective probe of quantities that would otherwise be accessible only through photon-coincidence measurements on the spontaneous pair, or only through fields whose partial coherence makes them awkward to describe within the coherent formalism developed so far.

The starting point for this shift is the isomorphism that here we make explicit. Consider the idler amplitude from SPDC, which appears in Eq.~\eqref{eq:spont_term}. Applying a transverse spatial mode filter that projects the signal photon onto the state defined by $v_F(\mathbf q_1)$ gives an idler amplitude given by
\begin{equation}
\mathcal A_{\mathrm{F}}(\boldsymbol{\rho}_2)
\propto \alpha
\iint
d\mathbf q_1
d\mathbf q_2
\,
e^{i\mathbf q_2\cdot\boldsymbol{\rho}_2}
\,
\Phi(\mathbf q_1,\mathbf q_2)
v_F(\mathbf q_1).
\label{eq:spont_filter}
\end{equation}
By direct comparison of the stimulated idler amplitude $\mathcal{A}_{\text{stim}}(\boldsymbol{\rho}_2)$ in Eq.~\eqref{eq:stimulated_term_amp} with the above equation, we see it has the same functional form as the amplitude $\mathcal A_{\mathrm{F}}$ with the conjugated seed angular spectrum $v_s^{*}(\mathbf{q}_1)$ standing in for the signal-photon projection. That is, 
any property of the biphoton state that survives this projection and is revealed in coincidence detection, is therefore encoded in a macroscopic, classically detectable beam, and any optical experiment on the biphoton has a StimPDC analogue in which one arm of the coincidence apparatus is replaced by a laser and a conventional photodetector. Three research programs have grown from this observation and are the subject of the present section: (i)~the classical characterization of two-photon states via stimulated emission tomography, which turns the isomorphism into a quantitative reconstruction protocol; (ii)~the experimental implementation of Klyshko's advanced-wave picture (AWP) with a real, macroscopic advanced wave, together with its application to nonlinear and undetected-light imaging; and (iii)~the transfer and conservation of partial-coherence properties, most prominently the twist phase of twisted Gaussian Schell-model (TGSM) beams, through nonlinear three-wave mixing, which generalizes the coherent transfer laws of Secs.~\ref{sec:fundamentals}--\ref{sec:vvb} to fields that are classical statistical mixtures rather than pure modes.

\subsection{Stimulated emission tomography}
\label{subsec:set}

The operational content of stimulated emission tomography, formulated by Liscidini and Sipe~\cite{Liscidini2013}, is a general relation between the average number of photon pairs emitted spontaneously in a given pair of modes and the number of photons produced when one of those modes is coherently seeded. In the low-gain, undepleted-pump regime, and for a seed mode matched to one of the modes populated by the spontaneous process, the average number of stimulated photons detected in the conjugate mode is proportional to the SPDC pair-generation probability density in those two modes:
\begin{equation}
n_{\text{stim}}(\eta,\omega_2) \;\approx\; |B_\nu(\omega_1)|^2\, n_{\text{spdc}}(\nu,\eta;\omega_1,\omega_2),
\label{eq:LS_ratio}
\end{equation}
with the proportionality constant set by the mean photon number $|B_\nu(\omega_1)|^2$ of the coherent seed. Because $|B_\nu|^2$ can be nine or more orders of magnitude larger than unity, the same joint spectral amplitude $|\phi_{\nu,\eta}(\omega_1,\omega_2)|^2$ that would take hours of coincidence counting to reconstruct at the single-photon level can be sampled in seconds with a classical spectrum analyzer. Crucially, SET does not require single-photon detectors, coincidence electronics, or ultra-low-noise cryogenics. It only requires a tunable, mode-selectable seed and a classical detector at the idler wavelength.

The proposal was rapidly turned into a practical characterization technique. Eckstein \emph{et al.}~\cite{eckstein2014} used difference-frequency generation to resolve the joint spectral density of an AlGaAs waveguide biphoton source with a resolution two orders of magnitude finer than had been possible via coincidence counting. Fang \emph{et al.}~\cite{Fang2014Optica} independently reached similar conclusions in fiber sources. Rozema \emph{et al.}~\cite{Rozema:15} then closed the conceptual loop by performing full polarization density-matrix tomography of a BBO ``sandwich'' entangled-photon source~\cite{Kwiat1999}, showing that SET yields fidelities indistinguishable from those of standard quantum-state tomography while being orders of magnitude faster. Follow-up work extended the technique to multidimensional and hyperentangled settings~\cite{Fang:16,Ciampini:19} and to strongly non-degenerate sources whose idler photons fall in wavelength ranges where single-photon detection is impractical~\cite{Yang2020SET}. In the low-gain regime, SET was shown to reconstruct not only the modulus but also the phase of the biphoton modal function~\cite{Keller2022}, closing the last conceptual gap with full state tomography.

The natural next step, transferring the SET philosophy from time--frequency to transverse-space degrees of freedom, was taken by Xu, Choudhary, and Boyd~\cite{Xu2024}. Their experiment seeds a type-II BBO crystal with Laguerre--Gauss modes $\mathrm{LG}_{0}^{\ell_s}$ prepared on a spatial light modulator, and measures the resulting distribution of LG modes in the stimulated idler by projective flattening followed by single-mode fiber coupling. The joint spatial mode distribution (JSMD) of the underlying biphoton state, defined by the amplitudes
\begin{equation}
C_{p_s,p_i}^{\ell_s,\ell_i} \;=\; \iint d\mathbf{k}_s\, d\mathbf{k}_i\, \Psi(\mathbf{k}_s,\mathbf{k}_i)\, \left[\mathrm{LG}_{p_s}^{\ell_s}(\mathbf{k}_s)\right]^{*}\!\left[\mathrm{LG}_{p_i}^{\ell_i}(\mathbf{k}_i)\right]^{*},
\label{eq:JSMD}
\end{equation}
is thereby reconstructed from purely classical measurements. For the collinear, thin-crystal limit and $p_s=p_i=0$, the coefficients collapse to the closed form~\cite{Xu2024}
\begin{equation}
|C_{0,0}^{\ell,-\ell}|^2 \;\propto\; \left(\frac{2\xi^2}{1+2\xi^2}\right)^{2|\ell|}, \qquad \xi \equiv w_p/w_{s,i},
\label{eq:Xu_LG}
\end{equation}
which the experiment reproduces quantitatively across a range of seed waists and $|\ell|$ up to several units. Beyond mere characterization, the same group has systematically exploited StimPDC as a controllable source of structured light: orthogonal spatial coding based on multiplexed seeds~\cite{Xu:23} and, more recently, engineered mode transfer in which the pump and seed transverse widths jointly determine the size and Rayleigh range of the stimulated beam across the LG, Hermite--Gauss, and angular bases~\cite{g22lcgr1}. Taken together, these works elevate SET from a spectral fingerprinting tool into a broadband, high-dimensional inspection technique for spatial entanglement, and they identify StimPDC as a natural bridge between quantum-state characterization and classical structured-light generation.

\subsection{Nonlinear imaging and Klyshko's advanced-wave picture}
\label{subsec:awp}

The advanced-wave picture, originally introduced by Klyshko and formalized by Belinskii and Klyshko~\cite{Belinskii1994AWP}, reinterprets coincidence imaging with SPDC as a classical-wave propagation problem: the detector that triggers a coincidence is replaced by an effective back-propagating (``advanced'') source that traverses the same optical elements as the conditioned photon, is reflected off the nonlinear crystal as if by a mirror whose amplitude profile and curvature is determined by the pump beam~\cite{Pittman1996}, and then propagates forward through the twin arm. Under the usual paraxial, thin-crystal, and mode-matching assumptions, and up to overall normalization, the coincidence distribution factorizes as
\begin{equation}
P_{\text{coinc}}(\phi_1,\phi_2) \;\propto\; I_{\text{AWP}}(\phi_2|\phi_1),
\label{eq:AWP_equivalence}
\end{equation}
where $I_{\text{AWP}}(\phi_2|\phi_1)$ is the intensity that a classical advanced field prepared in mode $\phi_1$ would produce at the detection plane of the twin arm. Projective measurements in the down-converted arms thereby map onto classical prepare-and-measure experiments with a single-mode advanced field. The AWP has for decades served as an intuitive design and alignment tool for SPDC-based imaging, ghost imaging, and Bell-type experiments. Aspden \emph{et al.}~\cite{Aspden2014AWP} provided the first full-field, camera-based verification of Eq.~\eqref{eq:AWP_equivalence} by directly comparing coincidence images with those produced by a back-illuminating diode, establishing the picture on a rigorous experimental footing.

The stimulated implementation of the AWP, developed by Arruda \emph{et al.}~\cite{Arruda2018AWP}, upgrades the back-propagating advanced wave from a heuristic auxiliary field into a real, macroscopic beam that can be seen on a standard CCD. The key observation is that Eq.~\eqref{eq:stimulated_term_amp} is isomorphic to the two-photon coincidence integral: when the seed is prepared as $v_s^{*}(\mathbf{q}_1)=\int d\mathbf{q}'_1\, H_1^{B}(\mathbf{q}_1,\mathbf{q}'_1)\,\varphi^{*}(\mathbf{q}'_1)$, that is, as the AWP field $\varphi^{*}$ propagated backwards through the transfer function $H_1^{B}$ of the signal-arm optics, the stimulated idler intensity becomes
\begin{align}
I_{\text{stim}}(\boldsymbol{\rho}_2) \;\propto\;  & \left|\int  d\mathbf{q}_1 d\mathbf{q}'_1\, d\mathbf{q}_2\, d\mathbf{q}'_2\, e^{i\mathbf{q}'_2\cdot\boldsymbol{\rho}_2}\,\varphi^{*}(\mathbf{q}'_1)\times \right. \nonumber \\ & \left. H_1^{B}(\mathbf{q}_1,\mathbf{q}'_1)
  v_p(\mathbf{q}_1+\mathbf{q}_2)\, H_2(\mathbf{q}_2,\mathbf{q}'_2)\right|^{2},
\label{eq:AWP_stim}
\end{align}
which mirrors the coincidence-count distribution obtained from Eq.~\eqref{eq:stimulated_term} in the SPDC scenario when the idler passes through optics described by transfer function $H_2$. When the pump is structured rather than a plane wave, the nonlinear crystal ceases to behave as an ordinary mirror and instead plays the role of an optically addressed spatial light modulator, imprinting the pump's amplitude and phase onto the advanced wave. Arruda \emph{et al.} exploited this to demonstrate, among other things, a lensless implementation of the fractional Fourier transform (FRFT) driven purely by the pump phase curvature, with the FRFT order
\begin{equation}
\cos\alpha \;=\; -\left(\frac{2z}{Z}+1\right)
\label{eq:FRFT_order}
\end{equation}
tuned by the position $Z$ of the pump waist relative to the crystal and the detection distance $z$, within the sign convention and propagation geometry of Ref.~\cite{Arruda2018AWP} (in which $Z<0$ for a pump convergent at the crystal). They further verified that a Hermite--Gauss pump generates a stimulated idler that reproduces the two-photon spatial correlations one would measure in coincidence. The result is a practical change in laboratory workflow: SPDC coincidence experiments can be designed, tested, and aligned using a StimPDC beam that is bright enough to be detected with a low-cost CCD camera, or even seen directly at some wavelengths, and only then switched to the photon-counting regime.

A conceptually related and equally influential development is the classical undetected-light imaging experiment of Cardoso \emph{et al.}~\cite{Cardoso2018ClassicalImaging}, which recasts the celebrated quantum experiment of Lemos \emph{et al.}~\cite{Lemos2014} (theoretically formalized by Lahiri \emph{et al.}~\cite{Lahiri2015}) in a fully classical setting. In the Cardoso scheme, a pump and a signal beam are reflected back through the same nonlinear crystal, generating two stimulated idler amplitudes that interfere at the detection plane. An object of complex transmittance $t(x,y)\, e^{i\vartheta(x,y)}$ placed only in the signal arm modulates the stimulated idler produced in the second passage, yielding an idler intensity
\begin{equation}
I(x,y) \;\propto\; \left|E\left(-\tfrac{x}{2},-\tfrac{y}{2};\omega_i\right)\right|^2 \left|1 + \big[t(-\tfrac{x}{2},-\tfrac{y}{2})\big]^{2}\, e^{i[2\vartheta-\phi]}\right|^{2},
\label{eq:Cardoso_intensity}
\end{equation}
whose visibility
\begin{equation}
V(x,y) \;=\; \frac{2\, t^2}{1+t^4}
\label{eq:Cardoso_visibility}
\end{equation}
depends nonlinearly on the object's amplitude transmittance. According to Eq.~\eqref{eq:Cardoso_visibility}, the local fringe visibility itself decreases as $t\to 0$ and reaches unity only in the transparent limit $t\to 1$. The practical advantage of the scheme for nearly opaque objects arises instead from the nonlinear dependence of the idler intensity on $t$, which for pure-amplitude objects is more conveniently characterized through the contrast
\begin{equation}
C(x,y) \;=\; \frac{I_0(x,y)}{I_0(x,y)+I_\pi(x,y)},
\label{eq:Cardoso_contrast}
\end{equation}
with $I_0$ and $I_\pi$ the intensities measured with a relative phase $\phi=0$ and $\phi=\pi$ between the two idler amplitudes. Because $C$ depends on $(1\pm t^2)^2$ rather than on $t$ alone, small variations of $t$ around zero translate into pronounced changes of $C$~\cite{Cardoso2018ClassicalImaging}, which is the mechanism that allows high-contrast imaging of nearly opaque targets---a regime in which linear schemes struggle. The Cardoso experiment clarifies the interpretational debate over the roles of induced coherence versus induced emission in the Lemos scheme, showing that the imaging effect does not require entanglement, only the phase and amplitude correlations transferred through StimPDC, and it establishes a practical, robust classical platform for nonlinear and undetected-light imaging. Its natural quantum counterpart has since been analyzed in detail, with resolution shown to be governed by the momentum correlation between detected and undetected photons rather than by the wavelength of the detected light~\cite{Fuenzalida2022Resolution}, placing the classical and quantum protocols on a common informational footing.

\subsection{Partial coherence, TGSM beams, and twist conservation}
\label{subsec:tgsm}

The third thread concerns the propagation of partially coherent, structured light through nonlinear three-wave mixing. The twisted Gaussian Schell-model beam introduced by Simon and Mukunda~\cite{Simon1993TGSM} is the canonical example of a partially coherent field carrying a nonseparable, chirality-bearing degree of freedom known as the twist phase. The cross-spectral density (CSD) of a TGSM beam at the waist reads
\begin{align}
W_{\text{TGSM}}(\mathbf{r},\mathbf{r}') \;=\; & A\, e^{-\frac{\mathbf{r}^2+\mathbf{r}'^{2}}{4w^{2}}}e^{-ik\frac{\mathbf{r}^{2}-\mathbf{r}'^{2}}{2R}} \times \nonumber \\
& e^{-\frac{|\mathbf{r}-\mathbf{r}'|^{2}}{2\delta^{2}}}\, e^{-ik\mu(xy'-yx')},
\label{eq:TGSM_CSD}
\end{align}
with $w$ the beam waist, $\delta$ the transverse coherence length, $R$ the radius of curvature, and $\mu$ the twist parameter. The positivity of the CSD imposes the fundamental bound
\begin{equation}
k|\mu| \;\le\; \frac{1}{\delta^2},
\label{eq:mu_bound}
\end{equation}
which shows immediately that the twist phase vanishes for coherent beams ($\delta\to\infty$). The twist is thus a genuine chirality of the statistical fluctuations of the field, not of any deterministic wavefront~\cite{Simon1993TGSM}. TGSM beams rotate about the propagation axis as they diffract and are known to be more resilient to atmospheric and oceanic turbulence than their untwisted counterparts, motivating an active experimental program in their synthesis and characterization~\cite{Canas2022TGSM}.

The interplay between TGSM beams and parametric processes was first analyzed in the spontaneous regime~\cite{Hutter20,Hutter21}, who showed that a partially coherent pump transfers a twisted-Gaussian structure to the biphoton, giving rise to what they termed twisted Gaussian biphotons that have an intricate entanglement structure. This built on and extended a broader line of previous work establishing how pump spatial coherence governs the transfer of position--momentum entanglement to the down-converted pair~\cite{Giese2018,Hug02019,Zhang2019PumpCoherence}: incoherent pumping preserves position correlations but destroys momentum entanglement, while continuously tuning the pump coherence smoothly interpolates between these regimes. In parallel, experimental work established robust methods for synthesizing and certifying TGSM beams from incoherent superpositions of coherent pseudo-modes displayed sequentially on an SLM~\cite{wang19,tian20,Canas2022TGSM}, thereby providing the input fields required by any StimPDC experiment with partially coherent light.

The theoretical and experimental description of partially coherent three-wave mixing in StimPDC was developed by dos~Santos \emph{et al.}~\cite{dosSantos2022PhaseConjugation}. The starting point is the observation that the stimulated idler field of Eq.~\eqref{eq:stimulated_term_amp} is bilinear in the pump and seed fields. Consequently, its CSD involves an ensemble average over products of pump and seed amplitudes,
\begin{equation}
W_i(\mathbf{r},\mathbf{r}') \;\propto\;
\langle E_p(\mathbf{r})\, E_s^{*}(\mathbf{r})\,
E_p^{*}(\mathbf{r}')\, E_s(\mathbf{r}')\rangle .
\label{eq:CSD_fourth_order}
\end{equation}
For statistically independent pump and seed ensembles, this average factorizes into the product of the individual second-order correlation functions, yielding
\begin{equation}
W_i(\mathbf{r},\mathbf{r}') =
W_p(\mathbf{r},\mathbf{r}')\, W_s^{*}(\mathbf{r},\mathbf{r}').
\label{eq:CSD_product}
\end{equation}
This reduction does not require the individual fields to obey Gaussian statistics. Rather, Gaussian moment factorization becomes relevant only when higher moments of the same stochastic field, or correlations between the pump and seed ensembles, are involved.

This is the natural CSD-level generalization of the amplitude product relation~\eqref{eq:product} established in Sec.~\ref{sec:theory}. Moreover, when both input fields are TGSM, one can show that Eq.~\eqref{eq:CSD_product} produces an idler that is itself a TGSM beam, with parameters
\begin{equation}
w_i^{-2} = w_p^{-2}+w_s^{-2}, \qquad \delta_i^{-2} = \delta_p^{-2}+\delta_s^{-2},
\label{eq:TGSM_widths}
\end{equation}
\begin{equation}
\frac{k_i}{R_i} = \frac{k_p}{R_p}-\frac{k_s}{R_s}, \qquad k_i\mu_i = k_p\mu_p - k_s\mu_s.
\label{eq:TGSM_conservation}
\end{equation}
The last of these is a conservation law for the twist phase, structurally identical to the well-known conservation of orbital angular momentum in coherent three-wave mixing already discussed in Sec.~\ref{sec:oam}, but now expressed at the level of the statistical fluctuations of the field. A particularly striking consequence follows when the pump is a coherent plane wave ($\mu_p=0$, $\delta_p\to\infty$, $w_p\to\infty$): the idler inherits the seed's amplitude and coherence envelope while its twist reverses in sign,
\begin{equation}
W_i(\mathbf{r},\mathbf{r}') \;\simeq\; W_s^{*}(\mathbf{r},\mathbf{r}'),
\label{eq:PC_conjugation}
\end{equation}
so that StimPDC realizes optical phase conjugation of a partially coherent beam~\cite{dosSantos2022PhaseConjugation}. This is the direct partially coherent analogue of the coherent phase-conjugation results reviewed in Sec.~\ref{sec:fundamentals} and, in the vectorial case, in Sec.~\ref{sec:vvb}. Because phase conjugation is the mechanism by which a distorted wavefront can be un-distorted upon back-propagation through the same aberrating medium, this result opens a route to wavefront correction of partially coherent beams propagating through turbulence, complementing the coherent-vortex protocols developed earlier.

The generality of the twist-conservation law was subsequently tested experimentally by dos~Santos \emph{et al.}~\cite{dosSantos2025TwistConservation}, who examined both second-harmonic generation (SHG) pumped by two independent TGSM beams and StimPDC seeded and pumped by independent TGSM beams. Introducing the normalized twist parameter $\tau \equiv \mu k \delta^2$ (bounded by unity in absolute value by Eq.~\eqref{eq:mu_bound}), the StimPDC conservation reads
\begin{equation}
\tau_i \;=\; \frac{\delta_s^2}{\delta_s^2+\delta_p^2}\,\tau_p \;-\; \frac{\delta_p^2}{\delta_s^2+\delta_p^2}\,\tau_s,
\label{eq:normalized_twist}
\end{equation}
while its SHG counterpart is a coherence-weighted sum of the two pump twists. Using the phase-randomized SLM pseudo-mode technique of Ref.~\cite{Canas2022TGSM} to synthesize independent TGSM pump and seed beams, and a double-slit interferometric method to read out $\mu_i$ from the tilt of the fringe pattern, the authors verified Eq.~\eqref{eq:normalized_twist} qualitatively across a wide range of $(\tau_p,\tau_s,\delta_p,\delta_s)$ combinations. Together, these results identify the twist phase as a robust statistical degree of freedom that survives nonlinear interaction even when the constituent fields are not pure modes but classical statistical mixtures, extending the mode-by-mode conservation laws of Sec.~\ref{sec:oam} to the partially coherent setting.

\section{StimPDC in the Quantum Regime}
\label{sec:quantum}

Everything discussed so far has treated the seed as a bright classical field, so
that the idler is macroscopically detectable and the whole apparatus behaves as a
piece of nonlinear classical optics that is conveniently cast using a quantum-optical description. The formalism of
Sec.~\ref{sec:theory}, however, never required $|\alpha|^{2}\gg 1$. As the seed is
attenuated toward the single-photon level, or as the idler is used not as a
signal but as a herald, the same interaction becomes a resource for genuinely
quantum operations. While not the focus of this review, this section briefly describes some of those uses. The key observation
is that the stimulated term of Eq.~\eqref{eq:intensity_split} does not switch off
when the seed becomes weak. It simply ceases to dominate, and one must then
either exploit it, post-select on it, or accept the spontaneous background as noise.

\subsection{Seeding at the single-photon level}
\label{subsec:singlephotonseed}

The obvious limit of StimPDC is a seed containing, on average, less than one
photon. The difficulty is equally obvious: the stimulated and spontaneous
contributions of Eq.~\eqref{eq:intensity_split} become comparable in intensity, and the
conjugated output is buried in the spontaneous background. The signal-to-noise
ratio that motivated the classical experiments of
Secs.~\ref{sec:fundamentals}--\ref{sec:vvb} is exactly what is lost.

A cascaded, heralded architecture circumvents this~\cite{Luck26}. A first source
produces a photon pair. The detection of one photon heralds the other, which is then
used as the seed of a second StimPDC stage, and the idler of that stage is
accepted only in coincidence with the herald. The coincidence with the herald removes
the spontaneous background of the second crystal, so that the phase-conjugation
relation of Eq.~\eqref{eq:product} can be applied to a true single photon. In the
transverse spatial degree of freedom this realizes single-photon orbital angular
momentum conjugation, $\ell \rightarrow -\ell$, at the cost of a count rate set
by the product of two down-conversion efficiencies.

\subsection{Stimulated emission, optimal cloning, and the universal-NOT gate}
\label{subsec:cloning}

A parametric amplifier seeded with a single photon in a definite polarization
state is the canonical physical realization of the optimal universal quantum
cloning machine. The amplified output in the seeded mode carries copies of the
input qubit with the optimal fidelity permitted by the no-cloning theorem, while
the orthogonally polarized output of the conjugate mode carries the optimal
approximation to the forbidden universal-NOT operation, that is, to the
orthogonal complement of an unknown qubit. Both were demonstrated in parametric
down-conversion~\cite{lamaslinares2002,demartini2002unot}, and the connection is
not an analogy: the anticloning output \emph{is} the stimulated idler of
Sec.~\ref{sec:theory} as applied to the entangled-photon sources in Sec.~\ref{subsec:vpconj}, and the optimality is a direct consequence of the
bosonic commutation relation used in Eq.~\eqref{eq:coherent_average}. The same physics constrains image amplification: the noise a phase-insensitive
amplifier must add is the counterpart of the fidelity bound on the optimal cloner,
and a phase-sensitive amplifier escapes both. This was first demonstrated in
$\chi^{(2)}$ parametric amplifiers~\cite{Choi1999PRL,Mosset2005PRL} and later
extended to hundreds of spatial modes simultaneously in
atomic vapor~\cite{Corzo2012Noiseless}.

The identification of the optimal universal NOT operation places the vector results of Sec.~\ref{sec:vvb} in a sharper
light. The Stokes-vector transformation realized by the two-crystal geometry,
$(S_1,S_2,S_3)\rightarrow(S_1,S_2,-S_3)$, is a conjugation with respect to a
particular basis, and, as noted in Sec.~\ref{subsec:vpconj}, one choice of
Bell-state source configuration produces instead the universal-NOT
map~\cite{deoliveira2020quantum,daSilva2021PRA}. The classical vector experiments
therefore implement, on a bright beam and with a CCD, precisely the single-qubit
operation that the single-photon experiments implement probabilistically. This is
a clear illustration of the isomorphism between
the spontaneous and stimulated regimes: the same interaction, read at two
different seed intensities, yields a quantum gate on one hand and a classical
mode converter on the other. 

\subsection{Heralded photon addition and non-Gaussian states}
\label{subsec:photonaddition}

The complementary use of StimPDC reverses the roles of the two output arms.
Rather than detecting the idler, one uses its detection as a herald: conditioned
on a click in the idler mode, a single photon has been added to whatever state
occupied the seed mode. The conditional operation implemented on the seed is,
to good approximation, the bare creation operator $\hat a^\dagger$, which is not
a unitary and not a Gaussian operation. Applied to a coherent state, first
proposed by Agarwal and Tara~\cite{AgarwalTara1991}, it produces the photon-added
coherent state $\hat a^\dagger \ket{\alpha}$, a state that interpolates
continuously between a Fock state and a coherent state as $|\alpha|$ is varied,
and whose Wigner function is negative for small $|\alpha|$ and positive for large
$|\alpha|$.

This was realized experimentally by Zavatta \emph{et al.}~\cite{zavatta2004},
who reconstructed the Wigner function of the single-photon-added coherent state
across the transition and thereby produced a direct, tomographic picture of the
quantum-to-classical boundary. Parigi \emph{et al.}~\cite{parigi2007} extended
the technique to the noncommutativity of addition and subtraction, and related
work established the corresponding thermal
cases~\cite{kiesel2007,kiesel2011}. In every case the enabling element is the
heralded StimPDC interaction described by Eq.~\eqref{eq:StimPDC_state}, now read
with the roles of the two channels exchanged.

The interest in these states is not only foundational. Photon-added and
photon-added-then-subtracted coherent states are non-Gaussian resources for
continuous-variable quantum key distribution, where the extremality theorems that
constrain coherent-state protocols no longer apply. They have been shown to
outperform coherent states in both post-selected homodyne and discrete-modulation
protocols, yielding higher key rates and longer maximum distances, with the
advantage growing as the channel becomes noisier, and to lend intrinsic robustness
against unambiguous state discrimination
attacks~\cite{Borelli2016QKD,Aguiar2022QKD,Srikara2020B92}. These protocols specify the states
but not their preparation. Heralded photon addition in a parametric interaction,
that is, StimPDC operated as in Ref.~\cite{zavatta2004}, is the established
physical route to them. Closing that loop, and in particular understanding how
the transverse mode structure of the heralding idler affects the purity of the
prepared state, is an open and practically relevant problem that the spatial
formalism of this review is well suited to partially address.

\subsection{Induced coherence and imaging with undetected photons}
\label{subsec:undetected}

In the induced-coherence
effect of Zou, Wang and Mandel~\cite{ZouWangMandel1991}, two down-conversion
sources are aligned so that the idler paths are indistinguishable. Then, interference appears in the signal arm and is controlled by an object placed in the idler
arm, even though no idler photon is ever detected. A more recent experiment by Lemos \emph{et
al.}~\cite{Lemos2014} turned this into full-field imaging with undetected
photons~\cite{Lahiri2015}.

The relation to the present review is direct and was made explicit by Cardoso
\emph{et al.}~\cite{Cardoso2018ClassicalImaging}, discussed in
Sec.~\ref{subsec:awp}: the same imaging effect survives when the ``undetected''
arm is seeded, that is, in the stimulated regime, which demonstrates that the
mechanism requires only the amplitude and phase correlations expressed by
Eq.~\eqref{eq:product} and not entanglement. The quantum version retains
advantages of its own, notably that its resolution is governed by the momentum
correlation between the detected and undetected photons rather than by the
wavelength of the detected light~\cite{Fuenzalida2022Resolution}, which is what
permits imaging at one wavelength while detecting at another. The classical and
quantum protocols are thus not competitors but two operating points of the same
interaction, and the practical question, which we take up in
Sec.~\ref{sec:outlook}, is which one to choose for a given imaging task. Broader
context for the quantum branch can be found in recent reviews of quantum
imaging~\cite{Defienne2024}.

\subsection{Quantum effects in the high-gain regime}
\label{subsec:highgain}
The preceding subsections approach the quantum regime from below, attenuating the
seed toward the vacuum. The opposite limit is equally instructive. Holding the
seed classical and increasing the parametric gain until the interaction can no
longer be truncated at first order, the output beams become macroscopically
populated and yet retain quantum correlations of their own. This regime lies
outside the scope announced in Sec.~\ref{sec:theory}, and work in this direction is extensive and consequential enough
to warrant a review of its own. It is nonetheless worth briefly summarizing here, because the questions addressed in that
regime are similar to those posed throughout this review, the
correspondence between the two regimes is instructive. Some of the
physical phenomena carry over directly, while others survive only in modified form, since the gain itself
begins to select the transverse modes that the seed alone selects at low gain. Establishing which results readily translate between
the two regimes, and which are properties of the perturbative limit rather than of
the interaction, is therefore of interest in both directions.
Table~\ref{tab:highgain} summarizes the current status.

\begingroup
\squeezetable
\begin{table}[t]
\caption{Status of the main results of this review in the high-gain regime.}
\label{tab:highgain}
\begin{ruledtabular}
\begin{tabular}{ll}
\C{0.24}{Result (low gain)} & \C{0.56}{High-gain counterpart} \\[4pt]
\hline
\C{0.24}{Phase conjugation [Sec.~\ref{subsec:asit}]}
& \C{0.56}{Survives; the conjugate beam and the amplified seed are quantum
correlated point by point across the transverse profile~\cite{Boyer2008Images}}
\\[4pt]
\C{0.24}{Brightness vs.\ cloning fidelity [Sec.~\ref{subsec:cloning}]}
& \C{0.56}{Phase-sensitive amplification is noiseless, and operates on many
spatial modes in
parallel~\cite{Choi1999PRL,Mosset2005PRL,Corzo2012Noiseless}} \\[4pt]
\C{0.24}{Transverse mode content [Sec.~\ref{sec:theory}]}
& \C{0.56}{Schmidt modes inherited from the low-gain biphoton, eigenvalues
redistributed toward low order~\cite{Sharapova2015PRA,Sharapova2020PRR}}
\\[4pt]
\C{0.24}{OAM conservation [Sec.~\ref{subsec:OAMconservation}]}
& \C{0.56}{Charge relation untested with a structured seed; correlations
between twin beams carrying OAM become delocalized~\cite{Marino2008OAM}}
\\[4pt]
\C{0.24}{Stimulated emission tomography [Sec.~\ref{subsec:set}]}
& \C{0.56}{Cascaded stimulated emission used to characterize high-gain
twin-beam sources~\cite{Triginer2020PRX}} \\[4pt]
\end{tabular}
\end{ruledtabular}
\end{table}
\endgroup

Two experimental traditions have developed this material. In $\chi^{(2)}$ media,
noiseless amplification of images in a traveling-wave optical parametric
amplifier was demonstrated by Choi \emph{et al.}~\cite{Choi1999PRL} and by Mosset
\emph{et al.}~\cite{Mosset2005PRL}, and sub-shot-noise spatial correlations were
observed in high-gain down-conversion~\cite{Jedrkiewicz2004PRL,Brambilla2004PRA}.
In $\chi^{(3)}$ four-wave mixing in hot alkali vapor, Boyer \emph{et
al.}~\cite{Boyer2008Images} seeded an amplifier with a probe carrying an image
and showed that the amplified probe and the generated conjugate form a pair of
entangled images. Because the traveling-wave geometry imposes no cavity mode
structure, a single pair of beams contains many independently correlated
subareas~\cite{Boyer2008Broadband,Embrey2015PRX}, whose size can be
controlled~\cite{Holtfrerich2016PRA} and used as an information
channel~\cite{Nirala2023SciAdv}. The two traditions differ less in physics than
in technique: resonantly enhanced four-wave mixing delivers large gain with a
continuous-wave pump, permitting shot-noise-limited homodyne detection, whereas
comparable gain in a bulk crystal requires pulsed pumping.

Read against the amplitude product relation~\eqref{eq:product}, the pattern in Table~\ref{tab:highgain} is
coherent. What survives is what follows from the structure of the interaction:
conjugation, and the mode basis itself, which remains that of the low-gain
biphoton~\cite{Sharapova2015PRA}. What changes is the weighting, since gain
favors the low-order Schmidt modes and thereby cleans the output mode at the
expense of the seed's control over it. The open question this review is best
placed to pose is what happens when those two mechanisms compete directly: a
structured seed selects one mode, gain narrowing selects another, and no
experiment has yet examined which wins, or in what sense the OAM and/or twist phase
survive.

\section{Applications and Outlook}
\label{sec:outlook}

The applications of StimPDC divide naturally according to which factor of the
amplitude product relation, Eq.~\eqref{eq:product}, is being exploited.

\emph{Wavefront correction and turbulence mitigation.} When the seed carries the
distortion, the idler carries its conjugate, and propagating the idler back
through the same aberrating medium undoes it. This is the nonlinear-optical
descendant of the phase-conjugating mirrors of the
1970s~\cite{zeldovich72}, with the added feature of wavelength conversion.
Recent work has applied it to free-space links, including all-optical turbulence
mitigation for quantum key distribution~\cite{aguilarcardoso2025turb} and
wavefront correction in StimPDC~\cite{Cardoso26}. The extension to partially
coherent beams follows from the twist-conservation results of
Sec.~\ref{subsec:tgsm}, where the phase conjugation acts at the level of the
cross-spectral density~\cite{dosSantos2022PhaseConjugation}, and is largely
unexplored experimentally.

\emph{All-optical spatial modulation at inaccessible wavelengths.} Because the
transfer function in Eq.~\eqref{eq:product} is written by the other beam rather
than by a physical device, StimPDC can shape light in spectral regions where
spatial light modulators do not exist. Demonstrations include topological-charge
readout by cross modulation~\cite{Santos2022}, engineered mode transfer across
the Laguerre--Gauss, Hermite--Gauss and angular bases~\cite{g22lcgr1}, and
random-phase filtering~\cite{Rocha2021Speckle}. A natural target is the
mid-infrared and the telecom band, where the idler can be structured by a visible
pump and read out, or conversely structured in the infrared and converted to a
wavelength where silicon cameras work.

\emph{Characterization of quantum sources.} Stimulated emission tomography
(Sec.~\ref{subsec:set}) reconstructs the joint spectral, polarization and now
spatial structure of a biphoton state with classical detectors and count rates
many orders of magnitude above coincidence counting. The advanced-wave picture
(Sec.~\ref{subsec:awp}) turns the same isomorphism into a laboratory workflow:
coincidence experiments can be designed and aligned with a bright stimulated beam
before being switched to the photon-counting regime. Both remain
under-exploited relative to their cost, which is essentially one additional
laser.

\emph{High-dimensional encoding and communication.} Orthogonal spatial coding
with multiplexed seeds~\cite{Xu:23} and vector-mode processing
(Sec.~\ref{sec:vvb}) point toward StimPDC as a mode-selective, wavelength-shifting
element in high-dimensional links. Suggested speckle-based encryption schemes~\cite{Rocha2021Speckle}, in which a scattering medium serves as a physical key applied to the idler,
is an appealing and so far untested variant.

\emph{Non-Gaussian state engineering.} Operated as a heralded photon adder
(Sec.~\ref{subsec:photonaddition}), the same interaction prepares the
non-Gaussian states that continuous-variable protocols call
for~\cite{Borelli2016QKD,Aguiar2022QKD}.

Beyond the applications sketched above, it is worth speculating about where
effort might most usefully be directed, since the gaps in the field are as
informative as its successes. Perhaps the most immediate of these is simply careful
bookkeeping. The literature is full of statements about brightness, but absolute conversion
efficiencies are rarely reported. Without them it is difficult to say when
StimPDC is genuinely preferable to the obvious alternative of a spatial light
modulator and a laser at the target wavelength. A systematic characterization of idler power as a function of
pump power, seed power, crystal length and mode overlap, reported in absolute
units for a few standard configurations, would be an unglamorous but genuinely
useful contribution.

A second direction concerns the boundaries of the approximations that make the
theory so compact. Equation~\eqref{eq:product} is exact only in the thin-crystal
limit, and although the correction is known formally as the convolution with the
phase-matching point-spread function in Eq.~\eqref{eq:pump_transfer_conv}, we are
not aware of a systematic experimental study of how the fidelity of image
transfer or phase conjugation degrades as that condition is relaxed. This is not
merely a matter of tidiness. If the degradation is well characterized it may also
be invertible, in which case deconvolution against a measured $\tilde\gamma$
would extend the usable spatial bandwidth of these techniques into the regime of
long, and therefore efficient, crystals, precisely where the efficiency question
raised above becomes pressing. A related question is what survives when the interaction can no longer be
truncated at first order. As summarized in Sec.~\ref{subsec:highgain}, the
unseeded case is largely understood, but the competition between a structured
seed and gain-induced mode selection has not been exhaustively examined, and the fate of OAM
and twist conservation under that competition remains open. Whether the amplitude product relation survives
in modified form, or whether mode competition selects a preferred spatial
structure independently of the seed, is an interesting direction for future study.

Finally, the entire body of work reviewed here has been performed with
bulk crystals in free space, at a moment when waveguide, periodically poled and
metasurface implementations of $\chi^{(2)}$ or $\chi^{(3)}$ interactions have become
routine. Confinement changes the problem qualitatively rather than
quantitatively: the transverse mode basis can become discrete and finite, the
thin-crystal condition is replaced by a modal overlap condition, and quantities
such as topological charge can lose their obvious meaning. It is not clear whether
the transfer laws survive this transition, are replaced by discrete selection
rules, or fail altogether, and the answer would determine whether the phase
conjugation and mode-conversion capabilities described in this review remain a
free-space laboratory technique or become an integrated photonic component. Moreover, understanding this behavior will enable the SET techniques discussed in Sec.~\ref{subsec:set} to be applied to modern quantum information experiments.

\section{Conclusion}
\label{sec:conclusion}
Stimulated parametric down-conversion occupies an unusual position in the context of classical and quantum optics. It is
described by the same Hamiltonian, the same phase-matching function and the same
two-photon amplitude as the spontaneous process that underpins the most important workhorse in photonic quantum
information, and yet its output is a bright, classically detectable beam. The
consequence, which we have tried to make explicit throughout, is that a single
relation, the amplitude product relation of Eq.~\eqref{eq:product}, connects a large body of results that are
usually presented separately: the image and coherence transfer experiments of the
1990s, the conservation and conjugation of orbital angular momentum, the
processing of vector vortex beams, the conservation of the twist phase of
partially coherent light, stimulated emission tomography, and the experimental
realization of Klyshko's advanced-wave picture. Each is a limit, a
generalization, or a re-reading of the statement that the stimulated idler is the
pump field multiplied by the conjugate of the seed.

Two directions seem to us the most promising. The first is instrumental: StimPDC
can give access to spatial modulation and wavefront control at wavelengths where
solid-state modulators do not exist, and this capability is still substantially
under-used. The second is methodological: because the stimulated and spontaneous
regimes are the same interaction at different seed intensities, almost any
experiment conceived for entangled photon pairs has a bright classical
counterpart that can be built, debugged and understood first. As the toolbox for structuring light continues to expand, we expect the StimPDC correspondence to become an increasingly valuable bridge between classical and quantum optics, providing both physical insight and practical guidance for the development of new structured-light techniques.

\begin{acknowledgements}
This work has been supported by the following Brazilian research agencies: Conselho Nacional de Desenvolvimento Cient\'{\i}fico e Tecnol\'ogico (CNPq - DOI 501100003593), Coordena\c c\~{a}o de Aperfei\c coamento de Pessoal de N\'\i vel Superior (CAPES DOI 501100002322), Funda\c c\~{a}o de Amparo \`{a} Pesquisa do Estado de Santa Catarina (FAPESC - DOI 501100005667), Instituto Nacional de Ci\^encia e Tecnologia de Infraestruturas Qu\^antica e
Nano para Aplica\c c\^oes Convergentes INCT-IQNano 406636/2022-2, Instituto Nacional de Ci\^encia e Tecnologia de Dispositivos Qu\^anticos (408783/2024-9). SPW and GS were supported by Fondo Nacional de Desarrollo Científico y Tecnológico (FONDECYT) Grant No.~1240746 and ANID Anillo Project ATE250003 -- Millennium Science Initiative Program -- ICN17\_012.
\end{acknowledgements}


\begin{thebibliography}{99}

\bibitem{ou90}
Z.~Y. Ou, L.~J. Wang, X.~Y. Zou, and L.~Mandel,
Phys. Rev. A \textbf{41}, 1597 (1990).

\bibitem{Ribeiro95}
P.~H. Souto~Ribeiro, S.~P\'adua, J.~C. Machado~da Silva, and G.~A. Barbosa,
Phys. Rev. A \textbf{51}, 1631 (1995).

\bibitem{monken1998transfer}
C.~H. Monken, P.~H. Souto~Ribeiro, and S.~P{\'a}dua,
Phys. Rev. A \textbf{57}, 3123 (1998).

\bibitem{fonseca1999quantum}
E.~J. da~Silva~Fonseca, P.~H. Souto~Ribeiro, S.~P{\'a}dua, and C.~H. Monken,
Phys. Rev. A \textbf{60}, 1530 (1999).

\bibitem{souto2001observation}
P.~H. Souto~Ribeiro, D.~P. Caetano, M.~P. Almeida, J.~A. Huguenin, B.~C. dos Santos, and A.~Z. Khoury,
Phys. Rev. Lett. \textbf{87}, 133602 (2001).

\bibitem{caetano2002conservation}
D.~P. Caetano, M.~P. Almeida, P.~H. Souto~Ribeiro, J.~A.~O. Huguenin, B.~C. dos Santos, and A.~Z. Khoury,
Phys. Rev. A \textbf{66}, 041801 (2002).

\bibitem{Oliveira21}
A.~G. de~Oliveira, G.~Santos, N.~Rubiano~da Silva, L.~J. Pereira, G.~B. Alves, A.~Z. Khoury, and P.~H. Souto~Ribeiro,
Phys. Rev. Appl. \textbf{16}, 044019 (2021).

\bibitem{Santos2022}
G.~H. dos Santos, A.~G. de~Oliveira, N.~Rubiano~da Silva, G.~Ca{\~n}as, E.~S. G{\'o}mez, S.~Joshi, Y.~Ismail, P.~H. Souto~Ribeiro, and S.~P. Walborn,
Nanophotonics \textbf{11}, 763 (2022).

\bibitem{deoliveira2020realtime}
A.~G. de~Oliveira, M.~F.~Z. Arruda, W.~C. Soares, S.~P. Walborn, R.~M. Gomes, R.~M. de~Ara{\'u}jo, and P.~H. Souto~Ribeiro,
ACS Photonics \textbf{7}, 249 (2020).

\bibitem{deoliveira2020quantum}
A.~G. de~Oliveira, N.~R. da~Silva, R.~M. de~Ara{\'u}jo, P.~H. Souto~Ribeiro, and S.~P. Walborn,
Phys. Rev. Applied \textbf{14}, 024048 (2020).

\bibitem{daSilva2021PRA}
N.~Rubiano~da Silva, A.~G. de~Oliveira, M.~F.~Z. Arruda, R.~Medeiros~de Ara\'{u}jo, W.~C. Soares, S.~P. Walborn, R.~M. Gomes, and P.~H. Souto~Ribeiro,
Phys. Rev. Applied \textbf{15}, 024039 (2021).

\bibitem{g22lcgr1}
A.~A. Aguilar-Cardoso, C.~Li, T.~J.~B. Luck, M.~F. Ferrer-Garcia, J.~Upham, J.~S. Lundeen, and R.~W. Boyd,
Phys. Rev. A \textbf{112}, 043541 (2025).

\bibitem{Xu:23}
Y.~Xu, S.~Tang, A.~N. Black, and R.~W. Boyd,
Opt. Express \textbf{31}, 42723 (2023).

\bibitem{aguilarcardoso2025turb}
A.~A. Aguilar-Cardoso, C.~Li, T.~J.~B. Luck, M.~F. Ferrer-Garcia, J.~Upham, J.~S. Lundeen, and R.~W. Boyd,
arXiv:2511.02066 (2025).

\bibitem{Liscidini2013}
M.~Liscidini and J.~E. Sipe,
Phys. Rev. Lett. \textbf{111}, 193602 (2013).

\bibitem{eckstein2014}
A.~Eckstein, G.~Boucher, A.~Lema{\^i}tre, P.~Filloux, I.~Favero, G.~Leo, J.~E. Sipe, M.~Liscidini, and S.~Ducci,
Laser Photon. Rev. \textbf{8}, L76 (2014).

\bibitem{Rozema:15}
L.~A. Rozema, C.~Wang, D.~H. Mahler, A.~Hayat, A.~M. Steinberg, J.~E. Sipe, and M.~Liscidini,
Optica \textbf{2}, 430 (2015).

\bibitem{Fang:16}
B.~Fang, M.~Liscidini, J.~E. Sipe, and V.~O. Lorenz,
Opt. Express \textbf{24}, 10013 (2016).

\bibitem{Ciampini:19}
M.~A. Ciampini, A.~Geraldi, V.~Cimini, C.~Macchiavello, J.~E. Sipe, M.~Liscidini, and P.~Mataloni,
Opt. Lett. \textbf{44}, 41 (2019).

\bibitem{Keller2022}
A.~Keller, A.~Z. Khoury, N.~Fabre, M.~Amanti, F.~Baboux, S.~Ducci, and P.~Milman,
Phys. Rev. A \textbf{106}, 063709 (2022).

\bibitem{Xu2024}
Y.~Xu, S.~Choudhary, and R.~W. Boyd,
Phys. Rev. Research \textbf{6}, L042047 (2024).

\bibitem{Luck26}
T.~J.~B. Luck, A.~A. Aguilar-Cardoso, C.~Li, M.~F. Ferrer-Garcia, J.~Upham, J.~S. Lundeen, and R.~W. Boyd,
J. Opt. Soc. Am. B \textbf{43}, A37 (2026).

\bibitem{lamaslinares2002}
A.~Lamas-Linares, C.~Simon, J.~C. Howell, and D.~Bouwmeester,
Science \textbf{296}, 712 (2002).

\bibitem{demartini2002unot}
F.~De~Martini, V.~Bu{\v{z}}ek, F.~Sciarrino, and C.~Sias,
Nature \textbf{419}, 815 (2002).

\bibitem{zavatta2004}
A.~Zavatta, S.~Viciani, and M.~Bellini,
Science \textbf{306}, 660 (2004).

\bibitem{parigi2007}
V.~Parigi, A.~Zavatta, M.~Kim, and M.~Bellini,
Science \textbf{317}, 1890 (2007).

\bibitem{kiesel2007}
T.~Kiesel, W.~Vogel, V.~Parigi, A.~Zavatta, and M.~Bellini,
Phys. Rev. A \textbf{78}, 021804(R) (2008).

\bibitem{kiesel2011}
T.~Kiesel, W.~Vogel, M.~Bellini, and A.~Zavatta,
Phys. Rev. A \textbf{83}, 032116 (2011).

\bibitem{ZouWangMandel1991}
X.~Y. Zou, L.~J. Wang, and L.~Mandel,
Phys. Rev. Lett. \textbf{67}, 318 (1991).

\bibitem{Lemos2014}
G.~B. Lemos, V.~Borish, G.~D. Cole, S.~Ramelow, R.~Lapkiewicz, and A.~Zeilinger,
Nature \textbf{512}, 409 (2014).

\bibitem{Defienne2024}
H.~Defienne, W.~P. Bowen, M.~Chekhova, G.~B. Lemos, D.~Oron, S.~Ramelow, N.~Treps, and D.~Faccio,
Nat. Photonics \textbf{18}, 1024 (2024).

\bibitem{Chekhova2016NLI}
M.~V. Chekhova and Z.~Y. Ou,
Adv. Opt. Photon. \textbf{8}, 104 (2016).

\bibitem{Wang1994}
L.~Wang,
\emph{Induced coherence in parametric down-conversion, both with and without induced emission},
Ph.D. thesis, University of Rochester (1994).

\bibitem{ribeiro1999}
P.~H. Souto~Ribeiro, S.~P{\'a}dua, and C.~H. Monken,
Phys. Rev. A \textbf{60}, 5074 (1999).

\bibitem{walborn2010}
S.~P. Walborn, C.~H. Monken, S.~P{\'a}dua, and P.~H. Souto~Ribeiro,
Phys. Rep. \textbf{495}, 87 (2010).

\bibitem{zeldovich72}
B.~Y. Zel'dovich, V.~I. Popovichev, V.~V. Ragul'skii, and F.~S. Faizullov,
JETP Lett. \textbf{15}, 109 (1972).

\bibitem{boyd_nlo}
R.~W. Boyd,
\emph{Nonlinear Optics}, 4th ed. (Academic Press, 2020).

\bibitem{Boyer2008Images}
V.~Boyer, A.~M. Marino, R.~C. Pooser, and P.~D. Lett,
Science \textbf{321}, 544 (2008).

\bibitem{Cardoso26}
A.~A. Aguilar-Cardoso, C.~Li, T.~J.~B. Luck, M.~F. Ferrer-Garcia, J.~Upham, J.~S. Lundeen, and R.~W. Boyd,
Optica \textbf{13}, 386 (2026).

\bibitem{ribeiro1994}
P.~H. Souto~Ribeiro, S.~P{\'a}dua, J.~C. Machado~da Silva, and G.~A. Barbosa,
Phys. Rev. A \textbf{49}, 4176 (1994).

\bibitem{santos2023}
G.~H. dos Santos, R.~C. Souza~Pimenta, R.~M. Gomes, S.~P. Walborn, and P.~H. Souto~Ribeiro,
Phys. Rev. Appl. \textbf{20}, 024007 (2023).

\bibitem{Hutter20}
L.~Hutter, G.~Lima, and S.~P. Walborn,
Phys. Rev. Lett. \textbf{125}, 193602 (2020).

\bibitem{Hutter21}
L.~Hutter, E.~S. Gomez, G.~Lima, and S.~P. Walborn,
AVS Quantum Sci. \textbf{3}, 031401 (2021).

\bibitem{arlt1999parametric}
J.~Arlt, K.~Dholakia, L.~Allen, and M.~J. Padgett,
Phys. Rev. A \textbf{59}, 3950 (1999).

\bibitem{tabosa1999optical}
J.~W.~R. Tabosa and D.~V. Petrov,
Phys. Rev. Lett. \textbf{83}, 4967 (1999).

\bibitem{dholakia1996second}
K.~Dholakia, N.~B. Simpson, M.~J. Padgett, and L.~Allen,
Phys. Rev. A \textbf{54}, R3742 (1996).

\bibitem{Allen1992}
L.~Allen, M.~W. Beijersbergen, R.~J.~C. Spreeuw, and J.~P. Woerdman,
Phys. Rev. A \textbf{45}, 8185 (1992).

\bibitem{padgett2000light}
M.~Padgett and L.~Allen,
Contemp. Phys. \textbf{41}, 275 (2000).

\bibitem{dosSantos2025TwistConservation}
G.~H. dos Santos, A.~L.~S. Santos~Junior, M.~Gil~de Oliveira, A.~C. Barbosa, B.~Pinheiro~da Silva, N.~Rubiano~da Silva, G.~Ca\~nas, S.~P. Walborn, A.~Z. Khoury, and P.~H. Souto~Ribeiro,
Phys. Rev. Appl. \textbf{24}, 034075 (2025).

\bibitem{de2019phase}
A.~G. de~Oliveira, M.~F.~Z. Arruda, W.~C. Soares, S.~P. Walborn, A.~Z. Khoury, A.~Kanaan, P.~H. Souto~Ribeiro, and R.~Medeiros~de Ara{\'u}jo,
Braz. J. Phys. \textbf{49}, 10 (2019).

\bibitem{rodrigues2018orbital}
R.~B. Rodrigues, J.~Gonzales, B.~Pinheiro~da Silva, J.~A.~O. Huguenin, M.~Martinelli, R.~Medeiros~de Ara{\'u}jo, C.~E.~R. Souza, and A.~Z. Khoury,
Opt. Lett. \textbf{43}, 2486 (2018).

\bibitem{Pereira17}
L.~J. Pereira, W.~T. Buono, D.~S. Tasca, K.~Dechoum, and A.~Z. Khoury,
Phys. Rev. A \textbf{96}, 053856 (2017).

\bibitem{buono2020chiral}
W.~Buono, A.~Santos, M.~Maia, L.~Pereira, D.~Tasca, K.~Dechoum, T.~Ruchon, and A.~Khoury,
Phys. Rev. A \textbf{101}, 043821 (2020).

\bibitem{Marino2008OAM}
A.~M. Marino, V.~Boyer, R.~C. Pooser, P.~D. Lett, K.~Lemons, and K.~M. Jones,
Phys. Rev. Lett. \textbf{101}, 093602 (2008).

\bibitem{Hickmann2010}
J.~M. Hickmann, E.~J.~S. Fonseca, W.~C. Soares, and S.~Ch\'avez-Cerda,
Phys. Rev. Lett. \textbf{105}, 053904 (2010).

\bibitem{Hickmann_2011}
J.~M. Hickmann, E.~J.~S. Fonseca, and A.~J. Jesus-Silva,
Europhys. Lett. \textbf{96}, 64006 (2011).

\bibitem{Rocha2021Speckle}
J.~C.~A. Rocha, D.~G. Pires, J.~G.~M.~N. Neto, A.~J. Jesus-Silva, N.~M. Litchinitser, and E.~J.~S. Fonseca,
Opt. Lett. \textbf{46}, 3905 (2021).

\bibitem{Borges10}
C.~V.~S. Borges, M.~Hor-Meyll, J.~A.~O. Huguenin, and A.~Z. Khoury,
Phys. Rev. A \textbf{82}, 033833 (2010).

\bibitem{Holleczek11}
A.~Holleczek, A.~Aiello, C.~Gabriel, C.~Marquardt, and G.~Leuchs,
Opt. Express \textbf{19}, 9714 (2011).

\bibitem{Milione2011}
G.~Milione, H.~I. Sztul, D.~A. Nolan, and R.~R. Alfano,
Phys. Rev. Lett. \textbf{107}, 053601 (2011).

\bibitem{Zhan09}
Q.~Zhan,
Adv. Opt. Photon. \textbf{1}, 1 (2009).

\bibitem{Souza2008}
C.~E.~R. Souza, C.~V.~S. Borges, A.~Z. Khoury, J.~A.~O. Huguenin, L.~Aolita, and S.~P. Walborn,
Phys. Rev. A \textbf{77}, 032345 (2008).

\bibitem{DAmbrosio2012}
V.~D'Ambrosio, E.~Nagali, S.~P. Walborn, L.~Aolita, S.~Slussarenko, L.~Marrucci, and F.~Sciarrino,
Nat. Commun. \textbf{3}, 961 (2012).

\bibitem{DAmbrosio2013}
V.~D'Ambrosio, N.~Spagnolo, L.~Del~Re, S.~Slussarenko, Y.~Li, L.~C. Kwek, L.~Marrucci, S.~P. Walborn, L.~Aolita, and F.~Sciarrino,
Nat. Commun. \textbf{4}, 2432 (2013).

\bibitem{Zhao2015}
N.~Zhao, X.~Li, and G.~Li,
Opt. Lasers Eng. \textbf{73}, 52 (2015).

\bibitem{Kwiat1999}
P.~G. Kwiat, E.~Waks, A.~G. White, I.~Appelbaum, and P.~H. Eberhard,
Phys. Rev. A \textbf{60}, R773 (1999).

\bibitem{Fang2014Optica}
B.~Fang, O.~Cohen, M.~Liscidini, J.~E. Sipe, and V.~O. Lorenz,
Optica \textbf{1}, 281 (2014).

\bibitem{Yang2020SET}
Y.~Yang, P.~Zhang, and X.-S. Ma,
J. Opt. Soc. Am. B \textbf{37}, 2071 (2020).

\bibitem{Belinskii1994AWP}
A.~V. Belinskii and D.~N. Klyshko,
J. Exp. Theor. Phys. \textbf{78}, 259 (1994).

\bibitem{Pittman1996}
T.~B. Pittman, D.~V. Strekalov, D.~N. Klyshko, M.~H. Rubin, A.~V. Sergienko, and Y.~H. Shih,
Phys. Rev. A \textbf{53}, 2804 (1996).

\bibitem{Aspden2014AWP}
R.~S. Aspden, D.~S. Tasca, A.~Forbes, R.~W. Boyd, and M.~J. Padgett,
J. Mod. Opt. \textbf{61}, 547 (2014).

\bibitem{Arruda2018AWP}
M.~F.~Z. Arruda, W.~C. Soares, S.~P. Walborn, D.~S. Tasca, A.~Kanaan, R.~Medeiros~de Ara\'ujo, and P.~H. Souto~Ribeiro,
Phys. Rev. A \textbf{98}, 023850 (2018).

\bibitem{Cardoso2018ClassicalImaging}
A.~C. Cardoso, L.~P. Berruezo, D.~F. \'Avila, G.~B. Lemos, W.~M. Pimenta, C.~H. Monken, P.~L. Saldanha, and S.~P\'adua,
Phys. Rev. A \textbf{97}, 033827 (2018).

\bibitem{Lahiri2015}
M.~Lahiri, R.~Lapkiewicz, G.~B. Lemos, and A.~Zeilinger,
Phys. Rev. A \textbf{92}, 013832 (2015).

\bibitem{Fuenzalida2022Resolution}
J.~Fuenzalida, A.~Hochrainer, G.~B. Lemos, E.~A. Ortega, R.~Lapkiewicz, M.~Lahiri, and A.~Zeilinger,
Quantum \textbf{6}, 646 (2022).

\bibitem{Simon1993TGSM}
R.~Simon and N.~Mukunda,
J. Opt. Soc. Am. A \textbf{10}, 95 (1993).

\bibitem{Canas2022TGSM}
G.~Ca{\~n}as, E.~S. G{\'o}mez, G.~H. dos Santos, A.~G. de~Oliveira, N.~Rubiano~da Silva, S.~Joshi, Y.~Ismail, P.~H.~S. Ribeiro, and S.~P. Walborn,
J. Opt. \textbf{24}, 094004 (2022).

\bibitem{Giese2018}
E.~Giese, R.~Fickler, W.~Zhang, L.~Chen, and R.~W. Boyd,
Phys. Scr. \textbf{93}, 084001 (2018).

\bibitem{Hug02019}
H.~Defienne and S.~Gigan,
Phys. Rev. A \textbf{99}, 053831 (2019).

\bibitem{Zhang2019PumpCoherence}
W.~Zhang, R.~Fickler, E.~Giese, L.~Chen, and R.~W. Boyd,
Opt. Express \textbf{27}, 20745 (2019).

\bibitem{wang19}
H.~Wang, X.~Peng, L.~Liu, F.~Wang, Y.~Cai, and S.~A. Ponomarenko,
Opt. Lett. \textbf{44}, 3709 (2019).

\bibitem{tian20}
C.~Tian, S.~Zhu, H.~Huang, Y.~Cai, and Z.~Li,
Opt. Lett. \textbf{45}, 5880 (2020).

\bibitem{dosSantos2022PhaseConjugation}
G.~H. dos Santos, A.~G. de~Oliveira, N.~Rubiano~da Silva, G.~Ca{\~n}as, E.~S. G{\'o}mez, S.~Joshi, Y.~Ismail, P.~H. Souto~Ribeiro, and S.~P. Walborn,
Nanophotonics \textbf{11}, 763 (2022).

\bibitem{Choi1999PRL}
S.-K. Choi, M.~Vasilyev, and P.~Kumar,
Phys. Rev. Lett. \textbf{83}, 1938 (1999).

\bibitem{Mosset2005PRL}
A.~Mosset, F.~Devaux, and E.~Lantz,
Phys. Rev. Lett. \textbf{94}, 223603 (2005).

\bibitem{Corzo2012Noiseless}
N.~V. Corzo, A.~M. Marino, K.~M. Jones, and P.~D. Lett,
Phys. Rev. Lett. \textbf{109}, 043602 (2012).

\bibitem{AgarwalTara1991}
G.~S. Agarwal and K.~Tara,
Phys. Rev. A \textbf{43}, 492 (1991).

\bibitem{Borelli2016QKD}
L.~F.~M. Borelli, L.~S. Aguiar, J.~A. Roversi, and A.~Vidiella-Barranco,
Quantum Inf. Process. \textbf{15}, 893 (2016).

\bibitem{Aguiar2022QKD}
L.~S. Aguiar, L.~F.~M. Borelli, J.~A. Roversi, and A.~Vidiella-Barranco,
Quantum Inf. Process. \textbf{21}, 304 (2022).

\bibitem{Srikara2020B92}
S.~Srikara, K.~Thapliyal, and A.~Pathak,
Quantum Inf. Process. \textbf{19}, 371 (2020).

\bibitem{Sharapova2015PRA}
P.~Sharapova, A.~M. P{\'e}rez, O.~V. Tikhonova, and M.~V. Chekhova,
Phys. Rev. A \textbf{91}, 043816 (2015).

\bibitem{Sharapova2020PRR}
P.~R. Sharapova, G.~Frascella, M.~Riabinin, A.~M. P{\'e}rez, O.~V. Tikhonova, S.~Lemieux, R.~W. Boyd, G.~Leuchs, and M.~V. Chekhova,
Phys. Rev. Research \textbf{2}, 013371 (2020).

\bibitem{Triginer2020PRX}
G.~Triginer, M.~D. Vidrighin, N.~Quesada, A.~Eckstein, M.~Moore, W.~S. Kolthammer, J.~E. Sipe, and I.~A. Walmsley,
Phys. Rev. X \textbf{10}, 031063 (2020).

\bibitem{Jedrkiewicz2004PRL}
O.~Jedrkiewicz, Y.-K. Jiang, E.~Brambilla, A.~Gatti, M.~Bache, L.~A. Lugiato, and P.~{Di Trapani},
Phys. Rev. Lett. \textbf{93}, 243601 (2004).

\bibitem{Brambilla2004PRA}
E.~Brambilla, A.~Gatti, M.~Bache, and L.~A. Lugiato,
Phys. Rev. A \textbf{69}, 023802 (2004).

\bibitem{Boyer2008Broadband}
V.~Boyer, A.~M. Marino, and P.~D. Lett,
Phys. Rev. Lett. \textbf{100}, 143601 (2008).

\bibitem{Embrey2015PRX}
C.~S. Embrey, M.~T. Turnbull, P.~G. Petrov, and V.~Boyer,
Phys. Rev. X \textbf{5}, 031004 (2015).

\bibitem{Holtfrerich2016PRA}
M.~W. Holtfrerich and A.~M. Marino,
Phys. Rev. A \textbf{93}, 063821 (2016).

\bibitem{Nirala2023SciAdv}
G.~Nirala, S.~T. Pradyumna, A.~Kumar, and A.~M. Marino,
Sci. Adv. \textbf{9}, eadf9161 (2023).

\end{thebibliography}
\end{document}